\documentclass[%
 reprint,
 amsmath,amssymb,
 aps,
]{revtex4-2}
\usepackage{graphicx}
\usepackage{dcolumn}
\usepackage{bm}

\begin{document}

\preprint{APS/123-QED}

\title{Modeling of uniflagellated bacterial locomotion in unbounded fluid and near a no-slip plane surface}
\author{Vahid Nourian}
 \email{vnourian@uwaterloo.ca}
\author{Henry Shum}%
 \email{henry.shum@uwaterloo.ca}
\affiliation{Department of Applied Mathematics, University of Waterloo, Waterloo, ON, N2L 3G1, Canada}%

\date{\today}

\begin{abstract}
The accumulation of swimming bacteria near surfaces may lead to biological processes such as biofilm formation and wound infection. Previous experimental observations of \textit{Vibrio alginolyticus} showed an interesting correlation between the bacterial entrapment near surfaces and the concentration of NaCl in the swimming medium. At higher concentrations of the ions, \textit{V.~alginolyticus} in the puller mode (with flagella in front of the body) tends to stay close to the surface whereas in the pusher mode (with flagella behind the body) it is more likely to escape from the surface. Motivated by these observations, we numerically investigate the locomotion of a uniflagellated model bacterium in unbounded fluid and near a planar surface. In our elastohydrodynamic model, the boundary integral technique and Kirchhoff rod model are employed respectively to calculate the hydrodynamic forces on the swimmer and model the elastic deformations of the flagellum consisting of a short, flexible hook and a long, relatively stiff filament. Our numerical results demonstrate that hydrodynamic interactions between the model bacterium and the solid wall cause the puller type to be attracted to the surface, whereas the pusher type is either repelled from or attracted to the surface depending on the flagellum and hook stiffness, the ion concentration (which determines the motor torque), and the cell body aspect ratio. We also show that large deformations of a very flexible hook can lead to an abrupt reorientation of the cell when the bacterium encounters a solid surface. These research findings can be used not only in understanding uniflagellated bacterial behavior but also in designing bacteria-mimicking micro-robots with biomedical and environmental monitoring applications.
\end{abstract}

\maketitle

\section{Introduction}
Bacterial habitats are usually confined by solid boundaries. The hydrodynamic interactions between the bacteria living in aqueous media and the boundaries could significantly affect the swimming properties of the bacteria.  Depending on the morphology and mode of motility of the species, the bacteria might be hydrodynamically trapped close to boundaries~\cite{wu2018,shum2010modelling,park2019flagellated}. Such entrapment of swimming bacteria near surfaces may facilitate some biological processes such as biofilm formation, which is a major problem in many industries and biomedical sectors~\cite{schierholz2001implant,conrad2012physics,bixler_biofouling_2012}. The search for practical solutions against biofilm formation requires a deep understanding of the pre-formation stages, including the hydrodynamic interactions between the bacteria and the contacted surfaces.

Monotrichous bacteria, such as \textit{Vibrio alginolyticus}, have a single flexible flagellum typically protruding from one pole of the cell body. The flagellar filament is approximately helical and is connected to the cell body via a very flexible and short hook. 

Peritrichous bacteria have multiple flagella distributed around the cell body. During steady swimming, the flagella generally form a single bundle behind the cell body~\cite{berg_bacteria_1973}. In these bacteria, the flexibility of the hook to bending motion allows it to act as a universal joint, transmitting torque from the motor to the filament even when the bundling causes the filament to be at an angle of 90$^\circ$ to the axis of the motor~\cite{brown_flagellar_2012}. In contrast, the role of flexibility of the hook and flagellar filament during steady swimming is less evident for monotrichous bacteria. Indeed, some theoretical studies that model thelocomotion of monotrichous bacteria assume that the hook and flagellum are rigid. With such models, Giacché et~al. and Shum et~al. demonstrated that bacteria can be trapped near no-slip surfaces and exhibit stable circular trajectories parallel to the surfaces~\cite{giacche2010hydrodynamic,shum2010modelling}. Their results indicated that the tendency of bacteria to swim close to surfaces and their stable distances from the surfaces strongly depend on the cell body's shape and the flagellum length. It was also shown that the bacteria are still attracted to a surface and exhibit stable periodic orbits when they swim between two parallel surfaces or at the corner of a rectangular channel~\cite{shum2015Parallel,shum2015rectangular}. Even though these results are qualitatively consistent with the experimental observations for steady swimming, the assumption of a rigid filament rotating about a fixed axis means that such models cannot determine how flexural rigidity or swimming speed affect the behavior of the bacterium near surfaces. Moreover, the bacterial motor is not always steady and the dynamic behaviour that occurs when the motor speed changes is not fully captured by rigid flagellum models.

For instance, previous experimental observations of \textit{V.~alginolyticus} show that the bacteria can swim in both forward (pushing) and backward (pulling) directions by switching the direction of the bacterial motor rotation while maintaining the same handedness of the helical flagellum~\cite{homma_chemotactic_1996}. The backward swimming speed is approximately 40\% greater than the forward swimming speed~\cite{magariyama_difference_2001}. Wu et~al.~\cite{wu2018} found that \textit{V.~alginolyticus} can be entrapped near a surface in both puller and pusher modes and that the entrapment behavior strongly depends on the cells' swimming speeds; the cell concentration near surfaces decreases with swimming speed for pushers and increases with swimming speed for pullers. 

It has also been observed that the cell body of \textit{V.~alginolyticus} sometimes abruptly reorientates when the motor switches from backward to forward swimming, a phenomenon known as flicking~\cite{xie_bacterial_2011}. To study the impacts of hook deformation on monotrichous bacterial motility, Shum and Gaffney extended the model by assuming that a rigid flagellum is connected to a spheroidal rigid cell body via a flexible and naturally straight hook~\cite{shum2012effects}. Unlike the most common rigid-flagellum models in which the flagellum shape is described with an amplitude growing factor (see~\cite{ramia1993role,shum2010modelling, giacche2010hydrodynamic}), they assumed that the filament is purely helical and connects tangentially to the hook, which bends during swimming to align the axes of the cell body and helical flagellum. They found that steady swimming of the bacteria near a surface is very sensitive to hook rigidity and the shapes of the cell body and flagellum. Even far from surfaces, effective swimming is possible only within a bounded range of hook stiffnesses scaled by the applied motor torque and hook length. 

In particular, an instability occurs when the motor torque exceeds a threshold. The instability causes the hook to bend and the flagellum to be become misaligned with the cell body axis. Son et al.~\cite{son2013} showed that a buckling instability was indeed involved in the flick of \textit{V.~alginolyticus}. Other theoretical studies of the role of hook flexibility identified a transition from straight to helical swimming as the rigidity decreases~\cite{nguyen_buckling_2017,zou_helical_2021}. Jabbarzadeh and Fu~\cite{jabbarzadeh_dynamic_2018} showed with simulations that a buckling instability of the hook, in conjunction with a flexible filament, can lead to flicking motion similar to those observed experimentally. The instability of the hook is also important for bundle formation in peritrichous bacteria, as shown by Riley et al.~\cite{riley_swimming_2018}.

Park et~al. numerically studied a flexible helical filament driven by rotations at one end~\cite{park2017instabilities} and demonstrated three dynamical states for the filament: stable twirling, unstable whirling, and stable overwhirling. The state that emerges is determined by physical properties such as the rotation frequency and the stiffness of the filament. To model the flicking behavior observed in \textit{V.~alginolyticus}, they assumed that the flagellar motor reversal causes the hook to temporarily relax into an unloaded state with a lower bending modulus, thereby increasing its susceptibility to buckling. In a subsequent study, a spheroidal cell body was included in the model and critical thresholds for the hook bending moduli and the rotation frequency for buckling instabilities were reported~\cite{park2019locomotion}. By assuming that the hook is in a relaxed mode for a short period of time, they investigated the effects of the hook stiffness and the rotation frequency on the buckling angles. In another study, Park et~al. numerically simulated the locomotion of a uni-flagellated bacterium with a rigid cell body and a flexible flagellum close to a planar surface~\cite{park2019flagellated}. They studied the influences of geometrical parameters, rotation frequency, and flagellum stiffness on the swimming properties of the model bacterium near the surfaces.

The sensitivity of the bacterial motion to the stiffness of the hook and filament equivalently manifests as sensitivity to the motor torque driving the flagellum. The modelling studies summarized above assumed either a prescribed motor torque or a prescribed rotation rate of the flagellum. It is known, however, that the torque generated by the bacterial motor follows a characteristic trend with rotation speed~\cite{berg_torque_1993, li_low_2006}; approximately constant torque is produced from 0 Hz up to a plateau knee frequency beyond which the torque drops linearly to zero. For \textit{V.~alginolyticus}, the motor is driven by Na$^+$ ions and Sowa et~al.~\cite{sowa2003torque} demonstrated that the torque--speed curve is shifted with the concentration of NaCl in the swimming medium such that higher torques are produced at higher concentrations. In any case, the observed torque and motor rotation speed is determined by the intersection of the motor torque--speed curve and the load line, which describes the linear dependence of the hydrodynamic torque on the filament with the motor speed. The slope of the load line can change as the hook or filament deform; hence, the torque and motor speed are generally time dependent.

In the present study, we use an elastohydrodynamic model to simulate the locomotion of a uniflagellated bacterium with a flexible hook and flagellum. A Regularized Stokes Formulation (RSF)~\cite{cortez_method_2001,ainley2008method,olson2013modeling,park2019flagellated} is accompanied by a Boundary Element Method (BEM)~\cite{pozrikidis2002practical} to model the hydrodynamic interactions among the bacterium components and a no-slip wall. Furthermore, we assume that the flagellum and hook are inextensible and unshearable and follow a discretization of the Kirchhoff rod model~\cite{lim2008dynamics}. It is expected that the hook's stiffness, the rest configuration, and the form of connection between the filament and the cell body considerably affect bacterial behavior. These effects and their importance on bacterial locomotion are further investigated in the first part of our work. We shed light on different aspects of locomotion of \textit{V.~alginolyticus} in bulk fluid and near a planar surface. Instead of applying a constant torque or constant rotational frequency to the motor, we assume that the flagellar motor follows the torque--speed curve obtained experimentally for \textit{V.~alginolyticus} at three levels of NaCl concentrations (3, 10, and 50 mM). 

Entrapment of three different strains of \textit{V.~alginolyticus} near surfaces is experimentally studied by Wu et~al.~\cite{wu2018}. Investigating the behavior of \textit{V.~alginolyticus} near a surface, in either puller or pusher modes, and different levels of NaCl concentrations is another aim of the present numerical study. Finally, the importance of the hook and flagellum flexibility and the cell body aspect ratio in near-surface entrapment of the uniflagellated bacteria is investigated.

\section{Modelling and Methods}
\begin{figure}
\includegraphics[width=85mm]{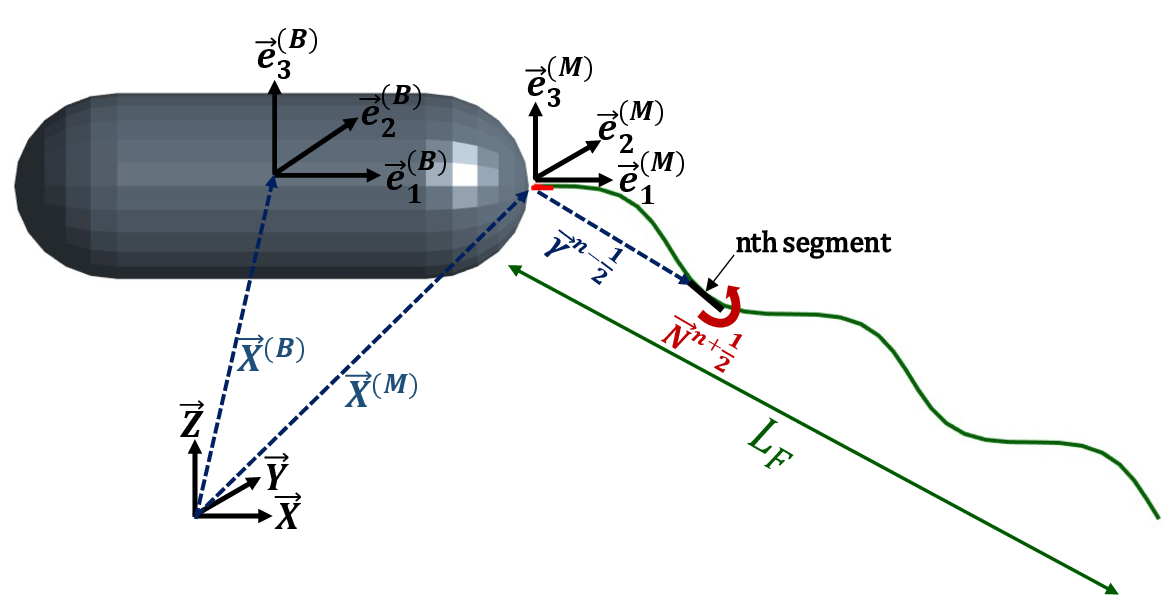}
\caption{\label{fig:Schematic} A schematic view of the model bacterium.}
\end{figure}

In this study, the uniflagellated model bacterium consists of a cell body, a flexible helical filament, and a very flexible straight or helical hook connecting the filament to a pole of the cell body. The dimensions of the  model bacterium follow the experimental measurements reported by Son et~al. for \textit{V.~alginolyticus}~\cite{son2013}. As illustrated in Fig.~\ref{fig:Schematic}, the shape of the cell body is a spherocylinder, i.e., a cylinder capped by a hemisphere at each end. A small gap is considered between the cell body and the hook to avoid singularities in the numerical scheme. The position and configuration of the model bacterium are described by three reference frames, namely, the motor-fixed frame $\left\{\vec{e}_1^\mathrm{(M)},\vec{e}_2^\mathrm{(M)},\vec{e}_3^\mathrm{(M)}\right\}$, body-fixed frame $\left\{\vec{e}_1^\mathrm{(B)},\vec{e}_2^\mathrm{(B)},\vec{e}_3^\mathrm{(B)}\right\}$, and global frame $\left\{\vec{X},\vec{Y},\vec{Z}\right\}$. We initialize all simulations with the hook and filament in their rest configurations. In the reference frame of the flagellar filament, the curve describing the shape of the filament in the rest configuration is given by:
{\small\begin{equation}
\vec{\Lambda}(\xi)=\xi\vec{e}_1^\mathrm{(F)}
+\Xi(\xi)\left[\cos{\left(\frac{2\pi}{p}\xi\right)}\vec{e}_2^\mathrm{(F)}+\sin{\left(\frac{2\pi}{p}\xi\right)}\vec{e}_3^\mathrm{(F)}\right].\label{equ:centerline}
\end{equation}}%
The variable $\xi$ parameterizes the distance along the axis of the right-handed helix with $0\leqslant\xi\leqslant L_\mathrm{F}$.  In this study, we compare two functional forms for $\Xi(\xi)$: a constant $\Xi(\xi) = a$, which describes a pure helical filament, and one with a growing factor $k_\mathrm{E}$ such that $\Xi(\xi)=a(1-e^{-(k_\mathrm{E}\xi)^2})$. In these equations, $a$ and $p$ represent the helix maximum amplitude and pitch, respectively. In the studied configurations, we use a hook of length $0.02l$ that is either straight or helical at rest and tangentially connected to the starting point of the filament. For the pure helical filament, the tangent constraint requires that the frame $\left\{\vec{e}_1^\mathrm{F)},\vec{e}_2^\mathrm{(F)},\vec{e}_3^\mathrm{(F)}\right\}$ rotate with respect to the motor fixed frame $\left\{\vec{e}_1^\mathrm{(M)},\vec{e}_2^\mathrm{(M)},\vec{e}_3^\mathrm{(M)}\right\}$. Therefore, the axis of the filament in the initial and rest configurations does not align with the cell axis. All the lengths in the model bacterium are non-dimensionalized by the averaged cell body radius $\overline{R}=0.81$\,\textmu{}m; this value is the radius of an equivalent sphere with the same volume as the cell body. Dimensions and mechanical properties of the model bacterium are given in Table~\ref{tab:physicalparameters}.  
 
In this study, motor torques, flagellum, and hook stiffnesses are non-dimensionalized with the maximum motor torque $T_\mathrm{Max}=3.8$\,pN\,\textmu{}m~\cite{sowa2003torque} in \textit{V.~alginolyticus}. We describe the relative stiffnesses of the hook and flagellum $k_{\varrho}(\varrho = H, F)$ as:
\begin{equation}
k_{\varrho}=\frac{(EI)_\mathrm{\varrho}}{T_\mathrm{Max}\overline{R}},
\end{equation}
where $E$ is Young’s modulus of the material and $I$ represents the moment of inertia of the cross sections of the hook and flagellum.

\begin{table*}
\caption{\label{tab:physicalparameters}
Dimensions and mechanical properties of the model bacterium
}
\centering
{\small{\begin{tabular}{|c|c|c|c|}
\hline \hline
\textrm{\textbf{Description}}&
\textrm{\textbf{Symbol}}&
\textrm{\textbf{Dimensionless value}}&
\textrm{\textbf{Dimensional value}}\\
 \hline
Cell body short radii & $R_2,R_3$ & 0.675 & 0.546\,\textmu{}m\\
Cell body long radius & $R_1$ & 2.5$R_2$ & 1.366\,\textmu{}m\\
Flagellum/Hook diameter & $d$ & 0.12 & 0.097\,\textmu{}m\\
Flagellum total length  & $l$ & 5.53 & 4.479\,\textmu{}m\\
Flagellum rest/initial pitch  & $p$ & 1.83 & 1.482\,\textmu{}m\\
Flagellum rest/initial amplitude  & $a$ & 0.172 & 0.138\,\textmu{}m\\
Hook length  & $l_\mathrm{H}$ & 0.02$l$ & 0.089\,\textmu{}m\\
Flagellum relative stiffness & $k_\mathrm{F}$ & 3.23 & -\\
(Flexural rigidity) & $(EI)_\mathrm{F}$ & - & (9.94\,pN\,\textmu{}m$^{2}$)\\
Hook relative stiffness & $k_\mathrm{H}$ & 0.125 & -\\
(Flexural rigidity) & $(EI)_\mathrm{H}$  & -   & (0.38\,pN\,\textmu{}m$^{2}$)\\
Repulsion strength of Lennard-Jones potential & $F_\mathrm{s}$ & 0.1 & 0.469\,pN\\
Cut-off distance of Lennard-Jones potential & $2^{1/6}\sigma$ & 0.2 & 0.162\,\textmu{}m\\
Number of segments on filament & $N_\mathrm{F}$ & 23 & -\\
Number of segments on hook & $N_\mathrm{H}$ & 2 & -\\
Number of triangular elements on the cell body & $N_{B}$ & 112 & -\\
Regularization parameter & $\epsilon$ & 0.5$d$ & 0.049\,\textmu{}m\\
Fluid viscosity & $\mu$ & 1 & 0.001\,N\,s\,m$^{-2}$\\
Fine time step & $\Delta t_\mathrm{fine}$ & $3.5\times 10^{-4}$ & $4.98\times 10^{-8}$\,s\\ 
Coarse time step & $\Delta t_\mathrm{coarse}$ & $3.5\times 10^{-2}$ & $4.98\times 10^{-6}$\,s\\
 \hline\hline
\end{tabular}}}
\end{table*}%
 
 \subsubsection{Hydrodynamic interactions}
 
The Reynolds number associated with \textit{V.~alginolyticus} motility in unbounded pure water is $\approx 10^{-4}$. In this Reynolds number, the inertia term in the Navier--Stokes equations is negligible and the flow field $\vec{u}$ is described by the incompressible Stokes equations
 {\begin{align}
-&\nabla p+\mu \Delta \vec{u}+\vec{F}_\mathrm{b}=\vec{0},\label{equ:NavierStokes}\\
&\nabla \cdot \vec{u}=0,\label{equ:Continuity} 
\end{align}}%
where $p$ is the fluid pressure, $\mu$ is the fluid viscosity, and $\vec{F}_\mathrm{b}$ is the body force field. In our formulation, the surface of the spherocylindrical cell body $B$ is treated as a rigid no-slip boundary for the fluid domain whereas the flagellar hook and filament exert regularized forces and torques along their centerlines, contributing to $\vec{F}_\mathrm{b}$. 

In a Lagrangian description, a point on the cell body surface $B$ is represented by $\vec{S}(\theta,\phi,t)$, where $\phi$ and $\theta$ are material coordinates on the cell's surface, and $t$ is time. The elastic flagellum (including both the hook and the filament sections), which rotates and deforms in time ($t$), is represented by a three-dimensional space curve $\Gamma(t)$, a point along which is denoted by $\vec{\gamma}(s,t)$, where $s$ is the arclength. Following these descriptions, the body force term $\vec{F}_\mathrm{b}$ due to the flagellum is expressed as:
{\begin{align}
\vec{F_\mathrm{b}}(\vec{x},t)=& 
\int_{\Gamma} {\vec{f}_\mathrm{F}(s,t)\psi_\epsilon(\vec{x}-\vec{\gamma}(s,t)) ds} \nonumber\\
+&\frac{1}{2}\nabla\times\int_{\Gamma} {\vec{n}(s,t)\Phi_\epsilon(\vec{x}-\vec{\gamma}(s,t))ds},\nonumber
\end{align}}%
where $\vec{f}_\mathrm{F}$ and $\vec{n}$ are respectively the force per unit length and torque per unit length applied to the fluid along the flagellum and the evaluation point $\vec{x}$ may be anywhere in the fluid including on the flagellum. The regularized stokes formulation~\cite{cortez2005method,olson2013modeling,park2019flagellated} is used here to reduce the slender flagellum to a one-dimensional distribution for computational efficiency, while approximately retaining a finite effective radius of the flagellum. The regularizing blob functions for the force and torque density are defined as
\begin{align}
\psi_\epsilon(\vec{x})&=\frac{15\epsilon^4}{8\pi ({\Vert \vec{x} \Vert}^2+\epsilon^2)^{7/2}},\label{equ:CutoffStokeslets}\\
\Phi_\epsilon(\vec{x})&=\frac{3\epsilon^2}{4\pi ({\Vert \vec{x} \Vert}^2+\epsilon^2)^{5/2}},\label{equ:CutoffRotlets}
\end{align}%
where we set $\epsilon=\frac{d}{2}$ to represent the effective flagellum radius. The velocity field due to the stress distribution $\vec{f}_\mathrm{B}$ on the cell body and the flagellum force and torque distributions in the presence of a no-slip plane wall at $z=0$ can be expressed in the form
\begin{align}
\vec{u}(\vec{x},t)= &\oint_{B} {\vec{U}_\mathrm{s}(\vec{f}_\mathrm{B},\vec{r}_\mathrm{B},\vec{\hat{r}}_\mathrm{B},0) dA}\label{equ:BIET}\nonumber\\
+&\int_{\Gamma}{\left[\vec{U}_\mathrm{s}(\vec{f}_\mathrm{F},\vec{r}_\mathrm{F},\vec{\hat{r}}_\mathrm{F},\epsilon)+\vec{U}_\mathrm{r}(\vec{n},\vec{r}_\mathrm{F},\vec{\hat{r}}_\mathrm{F},\epsilon)\right] ds},\nonumber\\
\end{align}%
 where $\vec{U}_\mathrm{s}$ and $\vec{U}_\mathrm{r}$ are respectively the velocities of the (regularized) stokeslet and rotlet including the image system for the no-slip wall. See Appendix~\ref{Regularized stokeslet} for the definitions of $\vec{U}_\mathrm{s}$ and $\vec{U}_\mathrm{r}$.
 
 The angular velocity field is obtained from the curl of the velocity field and is expressed in an analogous form, namely, 
 \begin{align}
\vec{w}(\vec{x},t)= &\oint_{B}{\vec{W}_\mathrm{s}(\vec{f}_\mathrm{B},\vec{r}_\mathrm{B},\vec{\hat{r}}_\mathrm{B},0) dA}\label{equ:BIEA}\nonumber\\
+&\int_{\Gamma}{\left[\vec{W}_\mathrm{s}(\vec{f}_\mathrm{F},\vec{r}_\mathrm{F},\vec{\hat{r}}_\mathrm{F},\epsilon)+\vec{W}_\mathrm{r}(\vec{n},\vec{r}_\mathrm{F},\vec{\hat{r}}_\mathrm{F},\epsilon)\right] ds}\nonumber\\
\end{align}%
where $\vec{W}_\mathrm{s}$ and $\vec{W}_\mathrm{r}$ are angular velocities of the regularized stokeslet and rotlet including the image system, defined in Appendix~\ref{Regularized stokeslet}.

\begin{figure}
\includegraphics[width=85mm]{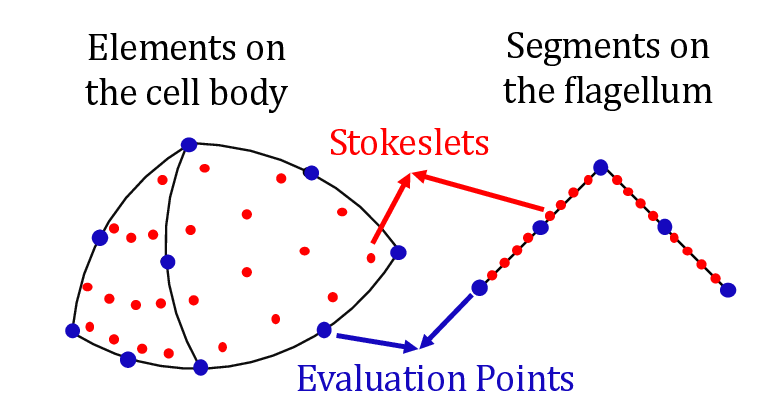}
\caption{\label{fig:Gauss} Distribution of Stokeslets and Rotlets over the curved triangular elements and along the connected straight rods. The blue evaluation points are placed at vertices, ends of segments, and the middle of edges and segments.}
\end{figure}

To evaluate the first integrals in. Eqs.~\ref{equ:BIET} and \ref{equ:BIEA}), the cell body surface $B$ is discretized into $N_{\mathrm{B}}$ curved triangular elements. For this purpose, six nodes are required to construct an element: three nodes are the vertices and a node at the middle of each edge (see Fig.~\ref{fig:Gauss}). We employ Gauss-Legendre quadrature method with 12 Gauss points to evaluate the integrals over the elements. Thus the surface of each element is mapped into a right-angle isosceles flat triangle before the integration. The stokeslets~(red points in Fig.~\ref{fig:Gauss}) are distributed according to Gauss-Legendre abscissas over each element, and their directions and magnitudes are approximated by using cardinal interpolation functions and interpolating the nodal force densities at the evaluation points (blue points)~\cite{pozrikidis2002practical,nourian_shum_2023}. Unlike the flagellum, which we treat as a one-dimensional object, the contribution from the cell body is a weakly singular surface integral that can be computed numerically without regularization, following the scheme presented by Pozrikidis~\cite{pozrikidis2002practical}.

The hook and flagellum are discretized into $N_\mathrm{H}$ and $N_\mathrm{F}$ connected equal-length straight segments. We use eight Gauss points on each segment to evaluate the integral. We also approximate the stokeslets/rotlets (red points in Fig.~\ref{fig:Gauss}) from the nodal force/torque densities at the evaluation points (blue points), by employing a second-order polynomial interpolation function.
In the following, we let $N_\mathrm{P_\mathrm{F}}=2(N_\mathrm{F}+N_\mathrm{H})+1$ denote the number of evaluation points on the flagellum and $N_\mathrm{P_\mathrm{B}}$ indicate the number of evaluation points on the cell body. We apply the presented scheme to Eqs.~(\ref{equ:BIET}, \ref{equ:BIEA}). By satisfying the no-slip boundary condition on the swimmer, a linear relationship between the translational and angular velocities (i.e. $\vec{u}_1,\hdots,\vec{u}_{N_\mathrm{P_\mathrm{B}}+N_\mathrm{P_\mathrm{F}}},\vec{w}_1,\hdots,\vec{w}_{N_\mathrm{P_\mathrm{F}}}$) and the nodal force and torque densities at the evaluation points (i.e., $\vec{f}_1,\hdots, \vec{f}_{N_\mathrm{P_\mathrm{B}}+N_\mathrm{P_\mathrm{F}}},\vec{n}_1,\hdots, \vec{n}_{N_\mathrm{P_\mathrm{F}}}$) is constructed.\\

\subsection{Kinematics}

Since the cell body in our model is rigid, the translational velocity of each node on the cell body's surface can be written in terms of the translational $\vec{U}^\mathrm{(B)}$ and angular $\vec{\Omega}^\mathrm{(B)}$ velocities of the cell body. Furthermore, the flagellar segments are only allowed to rotate with respect to the adjacent segments in the proposed model; thus the translational and angular velocities of each node on the flagellum can be expressed in terms of the relative angular velocities of the segments $\vec{\omega}_\mathrm{s}^\mathrm{n}$, $\vec{U}^\mathrm{(B)}$ and $\vec{\Omega}^\mathrm{(B)}$. Note that $\vec{\omega}_\mathrm{s}^\mathrm{n}$ represents  the angular velocity of the $n$th segment with respect to the $(n-1)$st one. This approach treats the filament as inextensible, which Jabbarzadeh \& Fu~\cite{jabbarzadeh_numerical_2020} showed could be much more computationally efficient than allowing extensibility while having an insignificant impact on the shape of the rotationally driven filament. 
 
The overall translational velocity of any given point $\vec{X}^E$ on the swimmer is written as:
{\small\begin{align}\label{equ:KinematicT}
&\vec{U}(\vec{X}^E) =\\
&\begin{cases}
\vec{U}^\mathrm{(B)}+\vec{\Omega}^\mathrm{(B)}\times\vec{X}^E, &\vec{X}^E\in\textrm{on cell body},\\\nonumber
\vec{U}^\mathrm{(B)}+\vec{\Omega}^\mathrm{(B)}\times\vec{X}^E+\displaystyle\sum_{n=1}^{m}{\vec{\omega}_\mathrm{s}^\mathrm{n}\times\vec{X}_\mathrm{rel}^\mathrm{n}},&\vec{X}^E\in m\text{th}\,\, \textrm{segment},\nonumber
\end{cases}
\end{align}}%
where
{\small\begin{equation}
\vec{X}_\mathrm{rel}^\mathrm{n}=\vec{X}^E-\vec{X}^\mathrm{(M)}-\vec{\gamma}^\mathrm{n-\frac{1}{2}}, \hspace{4mm}  n=1,...,N_\mathrm{H}+N_\mathrm{F};
\end{equation}}%
$\vec{\gamma}^\mathrm{n-\frac{1}{2}}$ is the position vector of the $n$th joint of the flagellum on the motor fixed frame, and $\vec{X}^\mathrm{(M)}$ denotes the position of the motor (flagellum base) on the cell body  (see Fig.~\ref{fig:Schematic}). The angular velocity of any given point $\vec{X}^E$ on the flagellum is the sum of the relative angular velocities of the preceding segments and the cell body's angular velocity as:
{\small\begin{equation}\label{equ:KinematicA}
\vec{w}(\vec{X}^E) =
\vec{\Omega}^{(B)}+\displaystyle\sum_{n=1}^{m}{\vec{\omega}_\mathrm{s}^\mathrm{n}},\hspace{7mm}\vec{X}^E\in m\text{th}\,\, \textrm{segment}.\\
\end{equation}}%
Finally, Eqs.~(\ref{equ:KinematicT},\ref{equ:KinematicA}) are written in form of a system of linear equations to represent the translational and angular velocities of all evaluation points in terms of the unknowns $\vec{\omega}_\mathrm{s}^\mathrm{n}$,  $\vec{U}^\mathrm{(B)}$ and $\vec{\Omega}^\mathrm{(B)}$~\cite{nourian_shum_2023}. 

\subsection{Elasticity}

In this study, the standard Kirchhoff rod model is employed to simulate the deformations of the hook and the flagellum. Since the stretching and shearing of the flagellum are negligible in comparison with the bending, we assume that the hook and the filament are inextensible, unshearable, and only allowed to bend and twist. The center line of the flagellum at the initial and rest configurations is represented by the space curve $\vec{\gamma}(s,0)$ (equivalent to $\vec{\Lambda}(\xi)$ in Eq.~\ref{equ:centerline} on the global frame). To describe the orientation of the material points in the cross-section of the flagellum, a right-handed orthonormal frame $\left\{\vec{D}_1(s,t),\vec{D}_2(s,t),\vec{D}_3(s,t)\right\}$ is introduced. For simplicity, it is assumed that $\vec{D}_3(s,t)$ is always tangent to the curve $\vec{\gamma}(s,t)$ i.e. $\vec{D}_3(s,t)=\vec{\gamma}'(s,t)$. Let $\vec{\kappa}(s,t)=(\kappa_1,\kappa_2,\kappa_3)$ denote the twist vector at point $s$ and time $t$. Then, following the linear theory of the elasticity, the internal moments $\vec{N}(s,t)$ transmitted along the flagellum can be estimated as~\cite{goriely1997nonlinear}: 

{\footnotesize\begin{align}
\vec{N}(s,t)=&EI\Big[(\kappa_1(s,t)-\hat{\kappa_1}(s))\vec{D}_1(s,t)+(\kappa_2(s,t)-\hat{\kappa_2}(s))\vec{D}_2(s,t) \nonumber\\
+&\Upsilon(\kappa_3(s,t)-\hat{\kappa_3}(s))\vec{D}_3(s,t)\Big],\label{equ:KirchhoffRod}
\end{align}}%
where $\vec{\hat{\kappa}}(s)$ represents the twist vector at the rest, and $\Upsilon=\frac{GJ}{EI}$ is the ratio of the twisting stiffness $GJ$ to the bending stiffness $EI$. Here, it is assumed that the flagellar filament and the hook are isotropic, homogeneous, and $\Upsilon=1$~\cite{park2019flagellated,park2019locomotion}.

\begin{figure}
\includegraphics[width=60mm]{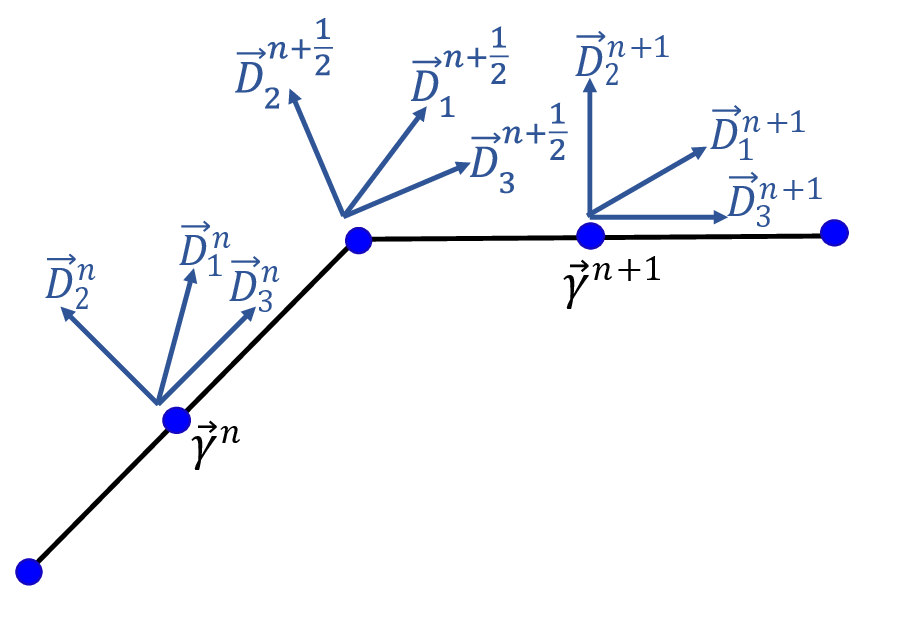}
\caption{\label{fig:Triad} Positions $\vec{\gamma}^{j}$ and triads $\{\vec{D}_{\hat{j}}^{j}$, $\hat{j} = 1,2,3\}$, at two successive segments $j=n$ and $j=n+1$. The triad at the joint denoted by$\{\vec{D}_{\hat{j}}^\mathrm{n+\frac{1}{2}}$, $\hat{j} = 1,2,3\}$, is obtained by interpolation of the triads of the neighboring segments.}
\end{figure}

\begin{figure}
\includegraphics[width=85mm]{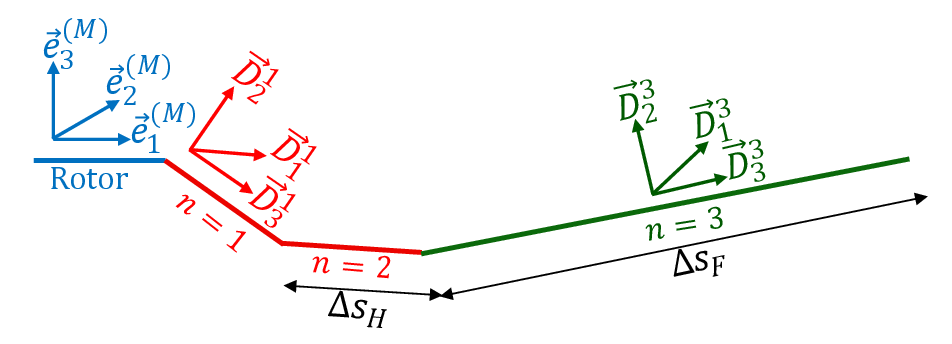}
\caption{\label{fig:Alignment}  The hook, connecting the rotor to the first segment of the filament, is discretized into $N_\mathrm{H}=2$ equal-length ($\Delta{s}_{\mathrm{H}}$) segments.}
\end{figure}

We discretize the hook and filament into $N_{\mathrm{H}}$  and $N_{\mathrm{F}}$ segments, respectively, by introducing uniform intervals $\Delta{s}_{\mathrm{H}}=l_{\mathrm{H}}/N_{\mathrm{H}}$ and $\Delta{s}_{\mathrm{F}}=(l-l_{\mathrm{H}})/N_{\mathrm{F}}$ of the Lagrangian variable $s$. The length of the hook is $l_\mathrm{H}$ and the total length of the flagellum (hook and filament) is $l$. In our model, the triads $\vec{D}_{\hat{i}}^\mathrm{n}$ ($n = 1, 2, ..., N_{\mathrm{H}}+N_{\mathrm{F}},\,\hat{i} = 1, 2, 3 $), which are updated over the time as the segments rotate, represents the orientation of the $n$th segment of the flagellum (see Figs.~\ref{fig:Schematic} and \ref{fig:Alignment}). The segment with index  $n=N_\mathrm{H}$ is the last segment of the hook and the index $n=N_\mathrm{H}+1$ represents the first segment of the filament. Since the segments on the hook are identical in length, the principal square root of the rotation matrix $M^\mathrm{n}$ that maps the triad $\{\vec{D}_{\hat{i}}^\mathrm{n}\}$ to the triad $\{\vec{D}_{\hat{i}}^\mathrm{n+1}\}$ is used to describe the orientation at the joint between neighboring segments, as shown in Fig.~\ref{fig:Triad}). A similar approach is used for the segments on the filament,
{\begin{align}
&M^\mathrm{n}=\sum_{{\hat{i}}=1}^3{\vec{D}_{\hat{i}}^\mathrm{n+1}(\vec{D}_{\hat{i}}^\mathrm{n})^T}, \label{equ:SquareRoot}\\
&\vec{D}_{\hat{i}}^\mathrm{n+\frac{1}{2}}=\sqrt{M^\mathrm{n}}\vec{D}_{\hat{i}}^\mathrm{n}.\label{equ:InterpolationTriad}
\end{align}}
By discretizing Eq.~(\ref{equ:KirchhoffRod}) and following the scheme used by Lim et~al.~\cite{lim2008dynamics}, the internal moments at the flagellum joints are estimated by:
{\begin{equation}
N_{\hat{i}}^\mathrm{n+\frac{1}{2}}=E_{\varrho}I_{\varrho}\left(\frac{\vec{D}_{\hat{j}}^\mathrm{n+1}-\vec{D}_{\hat{j}}^\mathrm{n}}{\Delta{s}_{\varrho}}\cdot\vec{D}_{\hat{k}}^\mathrm{n+\frac{1}{2}}-\hat{\kappa}_{\hat{i}}^\mathrm{n+\frac{1}{2}}\right),\label{equ:InternalMoment}
\end{equation}}
{\begin{equation}
\vec{N}^\mathrm{n+\frac{1}{2}}=\sum_{{\hat{i}}=1}^3{N_{\hat{i}}^\mathrm{n+\frac{1}{2}}\vec{D}_{\hat{i}}^\mathrm{n+\frac{1}{2}}},\label{equ:VectorInternalMoment}
\end{equation}}%
where subscript $\varrho=H, F$ distinguishes the hook from the filament ($\varrho=H$ for segment indices $n=1,...,N_\mathrm{H}$ and $\varrho=F$ for $n=N_\mathrm{H}+1,...,N_\mathrm{H}+N_\mathrm{F}-1$);  $({\hat{i}},{\hat{j}},{\hat{k}})$ is any cyclic permutation of $(1,2,3)$; $\vec{N}^\mathrm{n+\frac{1}{2}}$ is the internal moment transmitted from $n$th to $(n+1)$st segment; $\hat{\kappa}_{\hat{i}}^\mathrm{n+\frac{1}{2}}$ represents the twist vector's $\hat{i}$th component in the rest configuration, and $n=1,...,N_{\mathrm{H}}+N_{\mathrm{F}}$ is the segment number.

Above, $\vec{N}^\mathrm{\frac{1}{2}}$ denotes the internal moment transmitted from the rotor to the first segment of the hook. In the present scheme,  the magnitude and direction of $\vec{N}^\mathrm{\frac{1}{2}}$ are estimated by employing a sub-iterative method to satisfy the Kirchhoff rod model and impose the motor torque. See~\cite{nourian_shum_2023} for the details about this method.

\subsection{Steric repulsive force}

When the bacterium swims so close to the wall, the cell body and the flagellum are susceptible to touching the surface. Any collision causes the model to break, therefore we apply steric repulsive forces between the components and the wall at a short-range distance to keep them away.  A truncated Lennard-Jones potential is employed in our model to calculate the potential energy and then the corresponding repulsive force between the wall and the evaluation points on the swimmer~\cite{nguyen2018impacts,adhyapak2015zipping}. Specifically, we calculate the magnitude of the repulsive force between the $i$th evaluation point and the wall by finding the derivative of the Lennard-Jones potential ($U_{LJ}^{i}(h^i)$) with respect to the vertical distance of the point from the surface ($h^i$). Then, this force is applied to the point in the direction of the surface's normal vector.
{\begin{equation}\label{equ:Lennard-JonesPotential}
U_\mathrm{LJ}^{\mathrm{i}}(h^\mathrm{i})=\frac{F_\mathrm{s}\sigma}{6}\bigg[\left(\frac{\sigma}{h^\mathrm{i}}\right)^{12}-\left(\frac{\sigma}{h^\mathrm{i}}\right)^6\bigg]H(2^{1/6}\sigma-h^\mathrm{i}),
\end{equation}}
{\begin{equation}\label{equ:StericForce}
\vec{F}_\mathrm{rep}^{\mathrm{i}}=-\frac{dU_\mathrm{LJ}^{\mathrm{i}}(h^\mathrm{i})}{dh^{\mathrm{i}}}\vec{e}_3,
\end{equation}}%
where $i=1,\hdots,N_\mathrm{P_\mathrm{B}}+N_\mathrm{P_\mathrm{F}}$. We use a Heaviside step function $H$ to deactivate the repulsive force when the distance to the wall is more than the defined threshold $2^{1/6}\sigma$. In these equations, $\sigma$ is the cut-off distance, determines which kind of interaction occurs, and $F_\mathrm{s}$ is the repulsion strength. Our tests indicate that the magnitude of the repulsion strength does not have a significant impact on the locomotion of the bacterium as long as it is large enough to avoid collisions. For this reason, we choose a medium value for $F_\mathrm{s}$ which guarantees no collision. Moreover, we choose $2^{1/6}\sigma$ fairly greater than the effective radius of the flagellum ($\epsilon_\mathrm{F}$) to ensure that the flagellum does not touch the wall.\\

\subsection{Torque and force balance equations}
By neglecting the inertial and assuming that there is no gravity acting on the bacterium, the model bacterium is torque- and force-free~\cite {shum2010modelling,shum2019microswimmer}. In the total force balance Eq.~(\ref{equ:ForceBalance}), the sum of the steric repulsive forces and the integrals of viscous force densities over the elements (${B}^{n}$) and along the segments (${\Gamma}^{n}$) is set to be zero.
{\small\begin{equation}\label{equ:ForceBalance}
\sum_{n=1}^{N_\mathrm{B}}{\int_{B^{n}}{\vec{f}_\mathrm{B}dA_\mathrm{n}}}+\sum_{n=1}^{N_\mathrm{H}+N_\mathrm{F}}{\int_{\Gamma^{n}}{\vec{f}_\mathrm{F}ds_{n}}}+\sum_{i=1}^{N_{\mathrm{P_B}}+N_{\mathrm{P_F}}}{\vec{F}_\mathrm{rep}^i}=0. 
\end{equation}}%
 Likewise, the total torque about the center of the cell body is zero according to the torque balance equation
{\small\begin{align}\label{equ:TorqueBalance}
&\sum_{n=1}^{N_\mathrm{B}}{\int_{B^{n}}{(\vec{S}-\vec{X}^\mathrm{(B)})\times\vec{f}_\mathrm{B}dA_{n}}}+\sum_{n=1}^{N_\mathrm{H}+N_\mathrm{F}}{\int_{\Gamma^{n}}{\vec{n}ds_{n}}}\\\nonumber
+&\sum_{n=1}^{N_\mathrm{H}+N_\mathrm{F}}{\int_{\Gamma^{n}}{(\vec{X}^\mathrm{(M)}+\vec{\gamma}-\vec{X}^\mathrm{(B)})\times\vec{f}_\mathrm{F}ds_{n}}}+\sum_{i=1}^{N_{\mathrm{P_B}}+N_{\mathrm{P_F}}}{\vec{T}_\mathrm{rep}^i}=0,
\end{align}}%
where $\vec{T}_\mathrm{rep}^i$ is the torque applied to the center of the cell body due to the steric repulsive force at $i$th evaluation point. The first integral in Eq.~(\ref{equ:TorqueBalance}) represents the hydrodynamic torques induced by the viscous force densities on the cell body and the last two integrals describe the total torque densities on the flagellum and the torque induced by the viscous force densities, respectively.
To complete the system of the equations, we balance the viscous torques about each joint of the flagellum with the transmitted internal moment estimated by Eq.~(\ref{equ:VectorInternalMoment}):
{\small{\begin{align}\label{equ:JointsTorqueBalance}
\sum_{n=m}^{N_\mathrm{H}+N_\mathrm{F}}\Big[&\int_{\Gamma^{n}}{(\vec{\gamma}-\vec{\gamma}^{m-\frac{1}{2}})\times\vec{f}_\mathrm{F}ds_n}+\int_{\Gamma^{n}}{\vec{n}ds_n}\\\nonumber
&+\vec{T}_\mathrm{rep}^{2n}+\vec{T}_\mathrm{rep}^{2n+1}\Big]+\vec{N}^{m-\frac{1}{2}}=0,
\end{align}}}%
where $m=1,...,N_\mathrm{H}+N_\mathrm{F}$, and $\vec{T}_\mathrm{rep}^{2n}$, $\vec{T}_\mathrm{rep}^{2n}$ are the torques applied to the $m$th joint of the flagellum due to the steric repulsive forces at the middle and end of the $n$th segment. In fact, toque balance Eq.~(\ref{equ:JointsTorqueBalance}) is written for all the joints we have on the flagellum, thus $N_\mathrm{H}+N_\mathrm{F}$ equations are obtained in total. By employing the Gauss-Legendre quadrature method, Eqs.~(\ref{equ:ForceBalance}-\ref{equ:JointsTorqueBalance}) are expressed in terms of the nodal force and torque densities at the evaluation points (i.e., $\vec{f}_1,\hdots, \vec{f}_{N_\mathrm{P_\mathrm{B}}+N_\mathrm{P_\mathrm{F}}},\vec{n}_1,\hdots, \vec{n}_{N_\mathrm{P_\mathrm{F}}}$).

\subsection{Overview}
As already stated, the motor torque in our model is adjusted dynamically according to the rotation frequency of the flagellum. In this regard, the torque--speed curve in \textit{V.~alginolyticus} at three different concentrations of NaCl~\cite{sowa2003torque} are non-dimensionalized and employed here to apply a proper torque at a given concentration and motor frequency. If we split the torque--speed curve into two pieces, high torque-low speed, and low torque-high speed, and suppose that the relationship is linear in each piece, a piece-wise function can be constructed to relate the motor torque to its rotation frequency. The two lines intersect at the crossover point and the motor torque at a given frequency is the minimum of the two linear functions at that frequency. Specifically, for three levels of NaCl concentration, we model the torque-frequency relationships as:
\begin{align}\label{equ:Torque-Speed}
T_\mathrm{H}(\nu)&=\min{\{-1.203\nu+1,-25.197\nu+2.543}\},\nonumber\\ 
T_\mathrm{M}(\nu)&=\min{\{-1.691\nu+0.789,-24.572\nu+1.562}\},\nonumber\\
T_\mathrm{L}(\nu)&=\min{\{-1.071\nu+0.551,-33.079\nu+1.164}\}, 
\end{align}%
where $\nu$ represents the dimensionless motor frequency (revolutions per unit time), and $T_\mathrm{H}$, $T_\mathrm{M}$, and $T_\mathrm{L}$ denote the dimensionless motor torques at NaCl concentrations of $50$, $10$ and $3\,$mM, respectively. The equivalent experimental data were unavailable for motors running in reverse. For these cases, we assume that the torque--speed relationships are the same but with opposite signs. 

As shown in Fig.~\ref{fig:Alignment}, the axial direction $\vec{e}_1^\mathrm{(M)}$ indicates the rotor orientation, and therefore is fixed with respect to the cell body, whereas the orientations $\vec{e}_2^\mathrm{(M)},\vec{e}_3^\mathrm{(M)}$ change in time with the rotation of the rotor. To actuate the flagellum complex, the projection of the internal moment (at the joint connecting the hook to the rotor) onto $\vec{e}_1^\mathrm{(M)}$  sets to be equal to the motor torque at each time step, i.e.:
\begin{equation}
\vec{N}^\mathrm{\frac{1}{2}}\cdot\vec{e}_1^\mathrm{(M)}=T_\mathrm{i}(\nu), \hspace{3mm} i=L,M,H. \label{equ:motortorque}
\end{equation}
In this equation, $\vec{e}_1^\mathrm{(M)}$ is known and $T_\mathrm{i}(\nu)$ is obtained from Eq.~(\ref{equ:Torque-Speed}). A sub-iterative method (see the details in~\cite{nourian_shum_2023}) is employed here to solve Eq.~(\ref{equ:motortorque}) for $\vec{N}^{\frac{1}{2}}$.

Eqs.~(\ref{equ:BIET}-\ref{equ:KinematicT}, \ref{equ:KinematicA}, \ref{equ:ForceBalance}-\ref{equ:JointsTorqueBalance}) together construct a system of linear equations in which the components of the nodal force and torque densities at the evaluation points, the angular velocities of the segments, and the angular and translational velocities of the cell body are unknowns.  In this study, \texttt{mldivide} solver in Matlab is employed to solve the equations and determine the unknowns. Based on the obtained velocities, the configuration of the model bacterium is updated accordingly.

Combining a Kirchhoff rod model with a boundary element method leads to a stiff set of ODEs. Using general implicit schemes to solve these equations is computationally expensive. Instead, an explicit multirate time integration scheme is employed in this study. The original scheme, suggested by Bouzarth et~al.~\cite{bouzarth2010multirate}, includes spectral deferred corrections that we do not implement in our study. In this approach, the angular and translational velocities of the cell body and the nodal force densities on the cell body are updated on a coarse time step $\Delta t_\mathrm{coarse}$ while the angular velocities of the flagellar segments and the nodal force and torque densities on the flagellum are updated on a fine time step $\Delta t_\mathrm{fine}$. This procedure considerably reduces the computational time~\cite{nourian_shum_2023}.

\section{Results}
\subsection{Unbounded fluid}
We first investigate the locomotion of the model bacterium in free space. We compare the puller and pusher modes in terms of swimming speed and then explore the impacts of the shape of the hook and the model for the hook-flagellum transition on the swimming properties of the model bacterium. Except where otherwise noted, all simulations use a naturally straight hook and pure helical shape for the flagellar filament.

\subsubsection{Swimming speed}
\label{subsec:swimmingspeed}
\begin{figure}
\includegraphics[width=90mm]{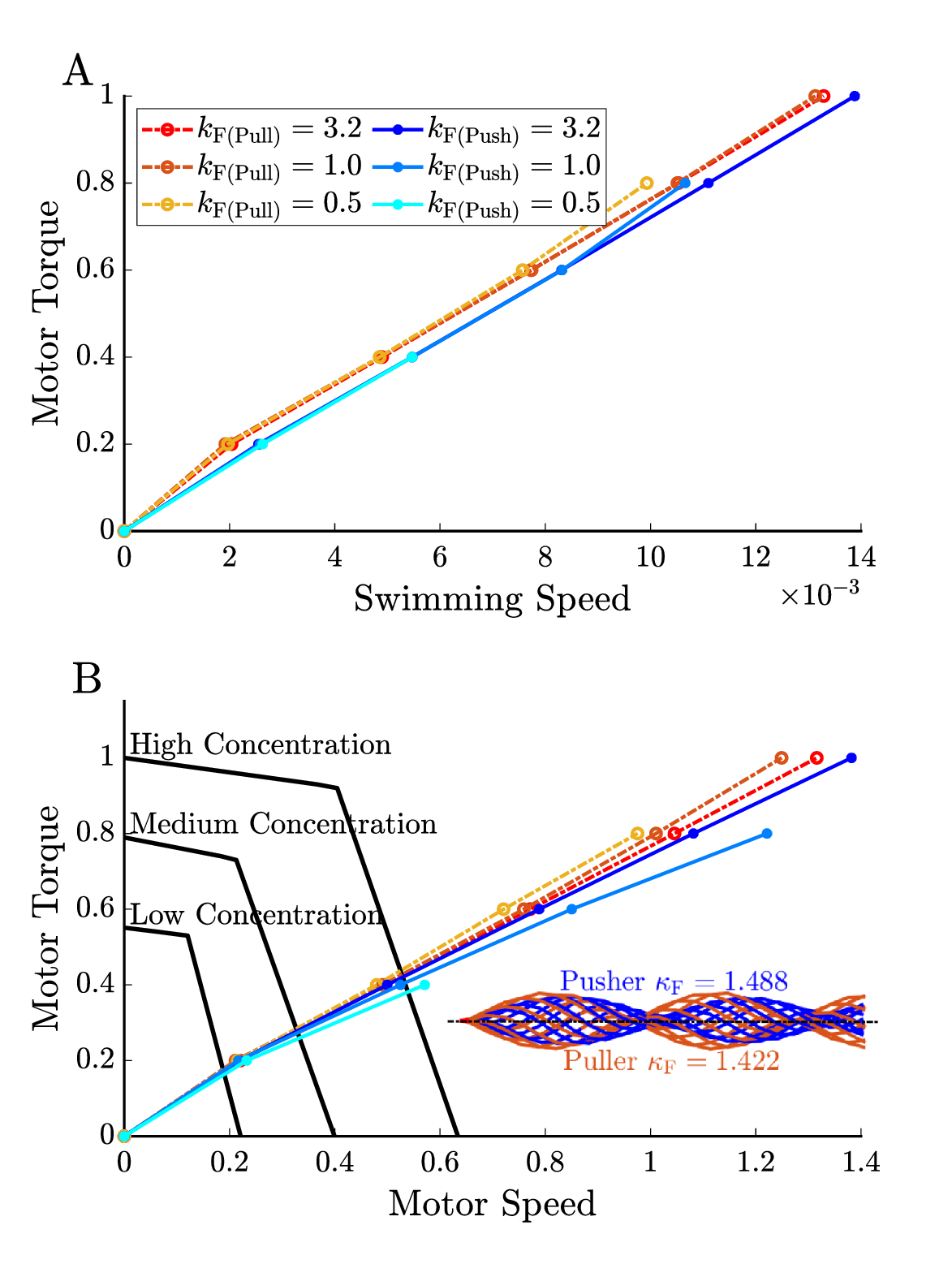}
\caption{\label{fig:SpeedTorque} A) The relationship between the swimming speed of the model bacterium and the motor torque in three filament stiffnesses and two swimming modes (puller/pusher). B) There is almost a linear relationship between the motor torque and motor speed. When the motor's torque dynamically changes with the motor speed, the intersection of the colored (load line) and the black curves (torque--speed curve) is an equilibrium point where the motor torque and speed converge to. The average filament's curvature $\kappa_\mathrm{F}=\sum_{j=N_\mathrm{H}+1}^{N_\mathrm{H}+N_\mathrm{F}-1}{[\vec{D}_3^{j+1}-\vec{D}_3^{j}]}/[\Delta{s}_\mathrm{F}(N_\mathrm{F}-1)]$ is slightly smaller in the puller flagellum than in the pusher one. The steady shapes of the puller and pusher flagella are shown over one period of the motor rotation on the motor-fixed frame.}
\end{figure}

To calculate the average swimming speed of the model bacterium in the unbounded space, we follow Higdon's formulation~\cite{higdon1979hydrodynamics} given as:
\begin{equation}\label{equ:AveragedSwimmingSpeed}
\overline{U}=\frac{(\vec{\Omega}^{(B)}-\vec{\omega}_s^0)\,\cdot\,\vec{U}^{(B)}}{||\vec{\Omega}^{(B)}-\vec{\omega}_s^0||}, 
\end{equation}
where $\vec{\omega}_s^0=2\pi\nu\vec{e}_1^\mathrm{(M)}$ is the angular velocity of the rotor. This formula is valid for motions that are constant in the motor-fixed frame, which would be the case once the flagellum reaches a steady shape and simply rotates about the motor axis. We apply constant motor torques, ranging from 0 to 1 in dimensionless units, and measure the steady swimming speeds in the puller and pusher modes. As shown in Fig.~\ref{fig:SpeedTorque}A, the swimming speed is approximately linear with the motor torque and it is slightly higher for the pusher than the puller at a given motor torque. Interestingly, for a given motor torque and mode, the swimming speed is almost independent of the flagellum stiffness $k_\mathrm{F}$ as long as the rotation of the flagellum is stable (when the flagellar rotation is unstable, changes in stiffness become important). Since the rotation of the flagellum becomes unstable at lower stiffnesses and/or higher motor torques, there are no data points for the puller with $k_\mathrm{F} = 0.5$ and $T = 1$, the pusher with $k_\mathrm{F}=1$, $T=1$, or the pusher with $k_\mathrm{F}=0.5$, $T\geq0.6$. It is worth stating that the pusher flagellum exhibits overwhirling rotation~\cite{park2017instabilities} and the puller flagellum bends toward the cell body and tends to wrap around it~\cite{park2022modeling} in the mentioned exceptional cases. 

For the same set of simulations with the motor torque varying from 0 to 1, we also track the rotation speeds of the flagellar motor ($2\pi\nu$) for the different stiffnesses of the flagellum and swimming modes. The obtained results indicate that the flagellum rotates slightly faster in the pusher than the puller at a given motor torque, as depicted in~Fig.~\ref{fig:SpeedTorque}B. Equivalently, for a fixed motor speed, the puller requires a larger motor torque than the pusher. A closer inspection of the filament shape indicates that the average curvature of the filament in the puller mode is slightly smaller than in the pusher mode. In other words, the puller filament has a slightly larger amplitude and/or pitch than the pusher one if we assume that the filament maintains an approximately helical shape during the rotation. This is consistent with simulation results of Park et al.~\cite{park2019locomotion}, who reported that the helical pitch and radius are smaller than their resting values during pushing motion and larger during pulling motion. 

Consequently, we find that the pusher flagellum rotates faster at a given motor torque because it has lower hydrodynamic resistance to axial rotations. The steady shapes of the rotating pusher and puller flagella on the body fixed-frame and over one period of the motor rotation are displayed in Fig.~\ref{fig:SpeedTorque}B. Unlike the swimming speed, the rotation frequency is affected by the stiffness of the filament. For pushers, increasing the flagellum stiffness decreases the motor speed at constant torque. For pullers, the opposite correlation is observed.

The slopes of the curves in Fig.~\ref{fig:SpeedTorque}B indicate the effective rotational drag coefficient of the flagellum depends on the flagellum stiffness. In contrast, the slopes of the curves in Fig.~\ref{fig:SpeedTorque}A are found to be relatively insensitive to the flagellum stiffness. This is an interesting observation because even though the flagellum deforms at higher torques and the rotational drag coefficient changes, the swimming speed maintains a linear relationship with the torque.

The torque--speed curves are characteristics of the flagellar motor and are independent of the properties of the attached flagellum. These curves at three concentrations of NaCl are depicted in Fig.~\ref{fig:SpeedTorque}B. The intersection of the flagellum's torque--speed curves (colored lines) and the motor's torque--speed curves [Eq.~(\ref{equ:Torque-Speed})] represents the motor's torque and speed during steady swimming. For a given NaCl concentration, Fig.~\ref{fig:SpeedTorque}B shows that pullers have a higher motor torque and lower motor speed than pushers with the same flagellum stiffness. 

In most of the remaining simulations, the instantaneous motor torque and frequency is computed at each time step to lie on the motor's torque--speed curve while simultaneously matching the hydrodynamic torque due to the rotation of the flagellum. In unbounded space, the shape of the flagellum generally stabilizes so the point of intersection converges to the equilibrium point presented in Fig.~\ref{fig:SpeedTorque}B. Any change in the motor load, due to the bacterium approaching a solid surface, for instance, alters the equilibrium point on the curve. Accounting for this behavior in our model is important because we can accurately study the locomotion of the bacteria in various swimming environments where changes in motor torque could be significant.

\subsubsection{Comparison of hook and flagellum shapes}

The hook, which acts as a  universal joint to transmit the torque from the rotor to the flagellar filament, could be straight or helical~\cite{son2013,shaikh2005partial} in the rest configuration. Motivated by this difference, we numerically study the impacts of the hook shape on the stable motor torque and the swimming speed in an unbounded fluid with a medium concentration of NaCl. 
 The actual molecular structure of the flagellar filament is uniform along its length, thus a purely  helical shape for the filament can be expected when it is stationary. In rigid models, the filament shape is usually described with an amplitude envelope growth rate $k_\mathrm{E}$ to align the flagellum's axis with the cell body's axis, as in Eq.~\ref{equ:centerline} with the second functional form for $\Xi$. Such an assumption is widely used in the literature but its effect on the swimming properties has not been quantitatively compared with the purely helical filament.

\begin{figure}
\includegraphics[width=88mm]{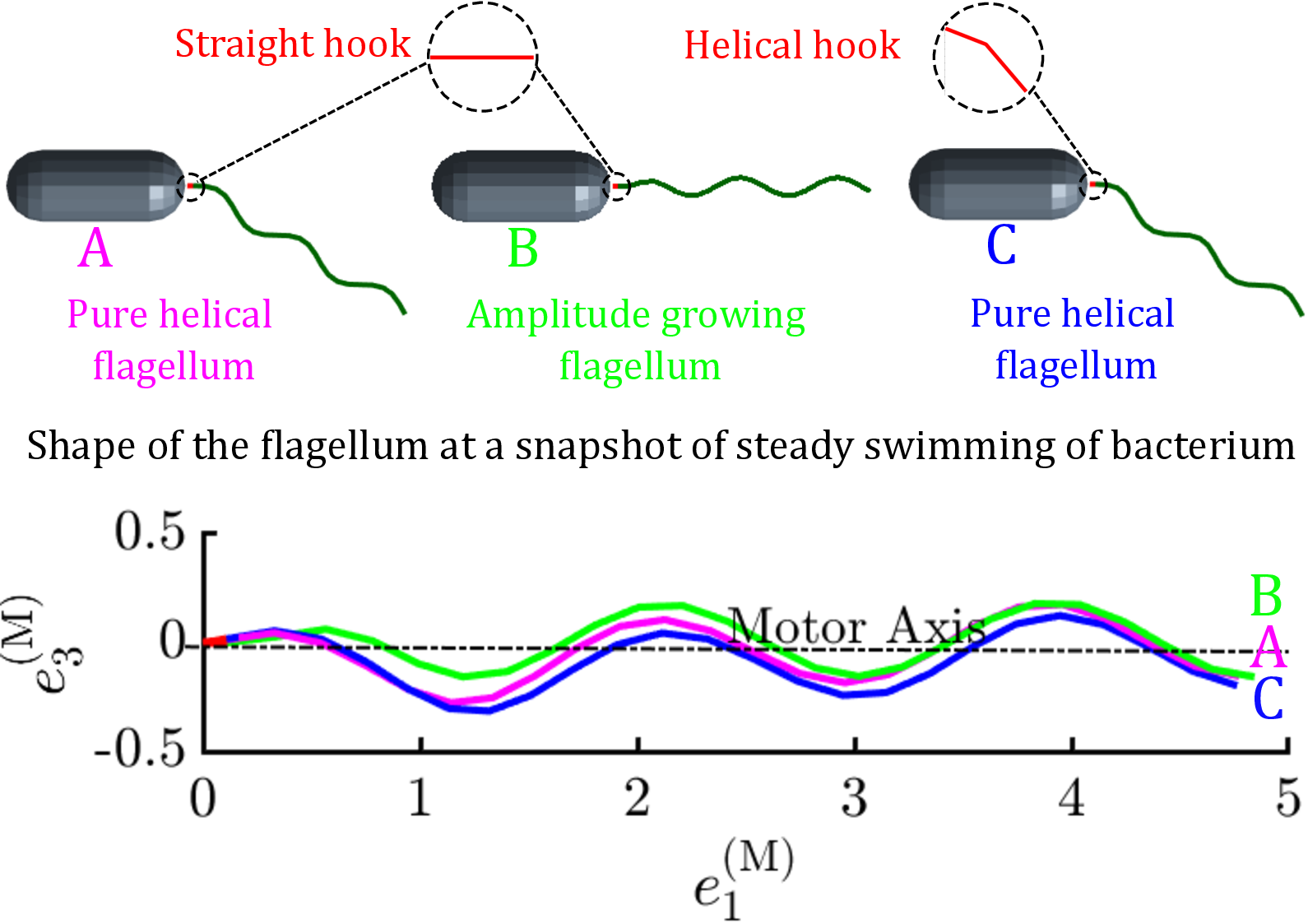}
\caption{\label{fig:UnboundedHook} Three different in-rest configurations are taken into account to compare the swimming properties of the model bacterium. The other physical properties of the model bacterium are stated in Tab.~\ref{tab:physicalparameters}. When the filament is purely helical, the axis of the filament does not align with the motor axis during the steady rotation. }
\end{figure}

To conduct this comparison, three model bacteria A, B, and C with different filament and hook configurations are taken into account. In the first configuration (A), the hook is straight and the rest shape of the filament is purely helical. In the second model bacterium (B), we assume that the hook is straight and the helical filament's rest shape is described using an amplitude growing rate $k_\mathrm{E}$. In the third one (C), the hook's rest shape is helical with the same properties as the filament, and the purely helical filament is tangentially connected to the hook, as shown in Fig.~\ref{fig:UnboundedHook}.
\begin{table}[b]
\caption{\label{tab:UnboundedHook}%
Comparing the transient time ($\Delta\,t_\mathrm{transient}$), steady-state swimming speed ($\overline{U}$), the steady motor torque ($T_\mathrm{steady}$) and motor speed ($2\pi\nu_\mathrm{steady}$) of the model bacterium with different hook and flagellum shapes in an unbounded fluid. The configuration of the model bacterium in cases A, B, and C are depicted in Fig.~\ref{fig:UnboundedHook}.}
\begin{ruledtabular}
\begin{tabular}{c|ccccc}
\textbf{Case} & mode & $\Delta\,t_\mathrm{transient}$ & $\overline{U}\times 10^3$ & $T_\mathrm{steady}$ & $2\pi\nu_\mathrm{steady}$\\
\hline
A & Pusher & 329  & 3.57  & 0.265 & 0.332\\
B & Pusher &  8   & 3.13  & 0.211 & 0.345\\
C & Pusher & 255  & 3.72  & 0.288 & 0.326\\
A & Puller & 279  & 3.31  & 0.265 & 0.332\\
B & Puller & 11   & 3.22  & 0.215 & 0.344\\
C & Puller & 207  & 3.14  & 0.287 & 0.326\\
\end{tabular}
\end{ruledtabular}
\end{table}%
The obtained results indicate that in configuration B, the model bacterium reaches steady-state quickly and its steady motor torque is the smallest among the cases in the puller and pusher modes (see Tab.~\ref{tab:UnboundedHook}). The transient time in our study is defined as the dimensionless time it takes for the speed calculated by \eqref{equ:AveragedSwimmingSpeed} to stabilize to within 2\% of its steady constant value. This represents the time required to relax to the steady swimming configuration from the initial conditions, which are at equilibrium in the absence of motor torques. In configurations A and C, the flagellum and the cell body's long axes are not initially aligned, thus longer transient times are obtained for these cases. After a few rotations of the flagellum, the angle between those axes eventually decreases and the swimming properties become steady. Closer inspection indicates that the flagellum curve remains stationary with respect to the motor-fixed frame during the steady state. As shown in Fig.~\ref{fig:UnboundedHook}, the axis of the flagellum does not align with the motor axis in configurations A and C, therefore off-axis rotation of the flagellum results in slightly different swimming properties.  

Even though the rotation speed of the pusher flagellum in configuration B is the highest among the cases, its swimming speed is the lowest, because the average amplitude of the filament in this configuration is smaller than in the others. In this regard, the pusher-type model bacterium with a helical hook has the highest swimming speed whereas it has the lowest motor speed. Moreover, the variation of the swimming speeds in the puller modes is small, but the differences in the steady motor torque are relatively considerable. This is an interesting observation because the model bacterium with configuration B swims as fast as the other model bacteria by applying a smaller torque to the flagellum.

\subsection{Near a surface}

Motivated by the behavior of \textit{V.~alginolyticus}, we mainly focus on the tendency of a uniflagellated model bacterium to a planar surface in this section. Specifically, the impacts of the swimming modes, NaCl concentration,  flagellar filament/hook stiffness, initial condition, and the cell body aspect ratio on the boundary accumulating behavior and escaping angles of the model bacterium are investigated.    

\subsubsection{Effects of concentration of sodium chloride}

\textit{V.~alginolyticus} utilizes a Na$^+$-driven flagellar motor to rotate the flagellum complex in the CW and CCW directions. The availability of sodium chloride in the swimming medium limits the performance of the motor. For this reason, the torque--speed relationship varies with the sodium chloride concentration, as expressed in Eq.~\ref{equ:Torque-Speed} and plotted in Fig.~\ref{fig:SpeedTorque}B. 
 \begin{figure}
\includegraphics[width=88mm]{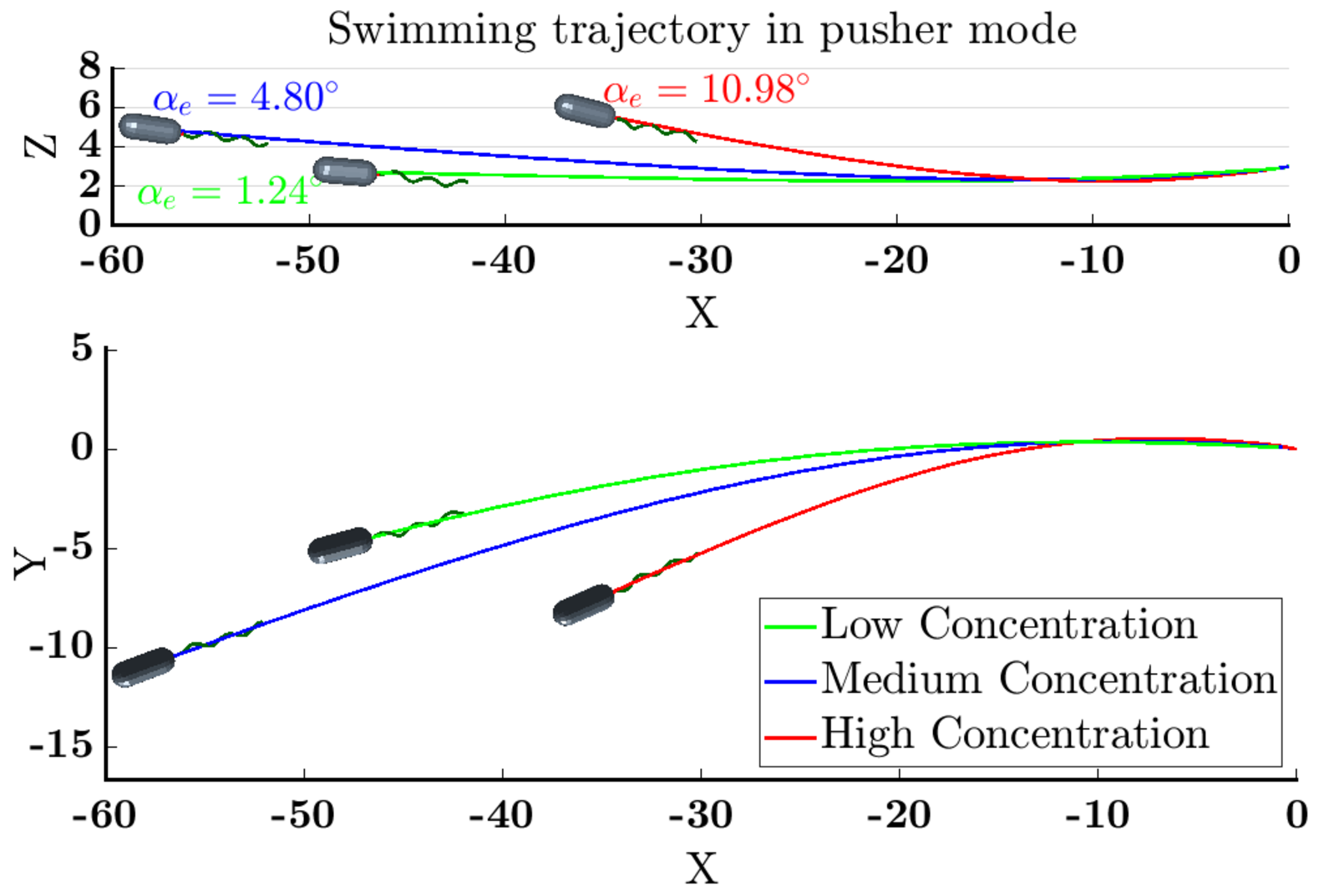}
\caption{\label{fig:PusherNearWall} Swimming trajectory of the model bacterium near a flat surface in pusher mode. In higher concentrations of NaCl, the bacterium immediately escapes from the surface with a relatively large angle ($\alpha_\mathrm{e}=10.98^\circ$). The escaping angle $\alpha_\mathrm{e}$ represents the angle between the swimming trajectory during the escaping state and the surface. The bacteria are initialized at $z=3$ and with the cell body tilted $15^\circ$ below the horizontal (towards the wall). Trajectories are shown up to times $T_\mathrm{s}=23000$, $T_\mathrm{s}=17000$, and $T_\mathrm{s}=7000$ for the low, medium, and high concentrations, respectively.}
\end{figure}
The swimming trajectories in Fig.~\ref{fig:PusherNearWall} demonstrate that the model bacterium in the pusher mode tends to escape from the surface at all three concentrations of NaCl. The escaping angle $\alpha_\mathrm{e}$ increases significantly with sodium concentration; at low concentrations, the bacterium swims almost parallel to the surface ($\alpha_\mathrm{e} \approx 1^\circ$). These results are consistent with the experimental observation of \textit{V.~alginolyticus} in which higher concentrations of cells are observed near the surface under lower concentrations of Na$^+$ ions in the swimming medium~\cite{wu2018}. This trend can be explained by the dipolar structure of the flow field generated by a pusher bacterium. When such a swimmer is approximately parallel to a wall, the image flow field due to the no-slip boundary pulls the swimmer towards the wall. The hydrodynamic attraction is strongest at the centre, around the hook, causing the hook to bend and the cell body to point away from the wall. The result is that the bacterium tends to swim away from the wall. At higher concentrations of NaCl, the hydrodynamic stresses are larger so the hook bends more. In our simulations, we found that the maximum angles between the cell body's long axis and the surface were $14.05^\circ$, $8.79^\circ$ and $7.01^\circ$, respectively, in decreasing order of NaCl concentrations.

 \begin{figure}
\includegraphics[width=80mm]{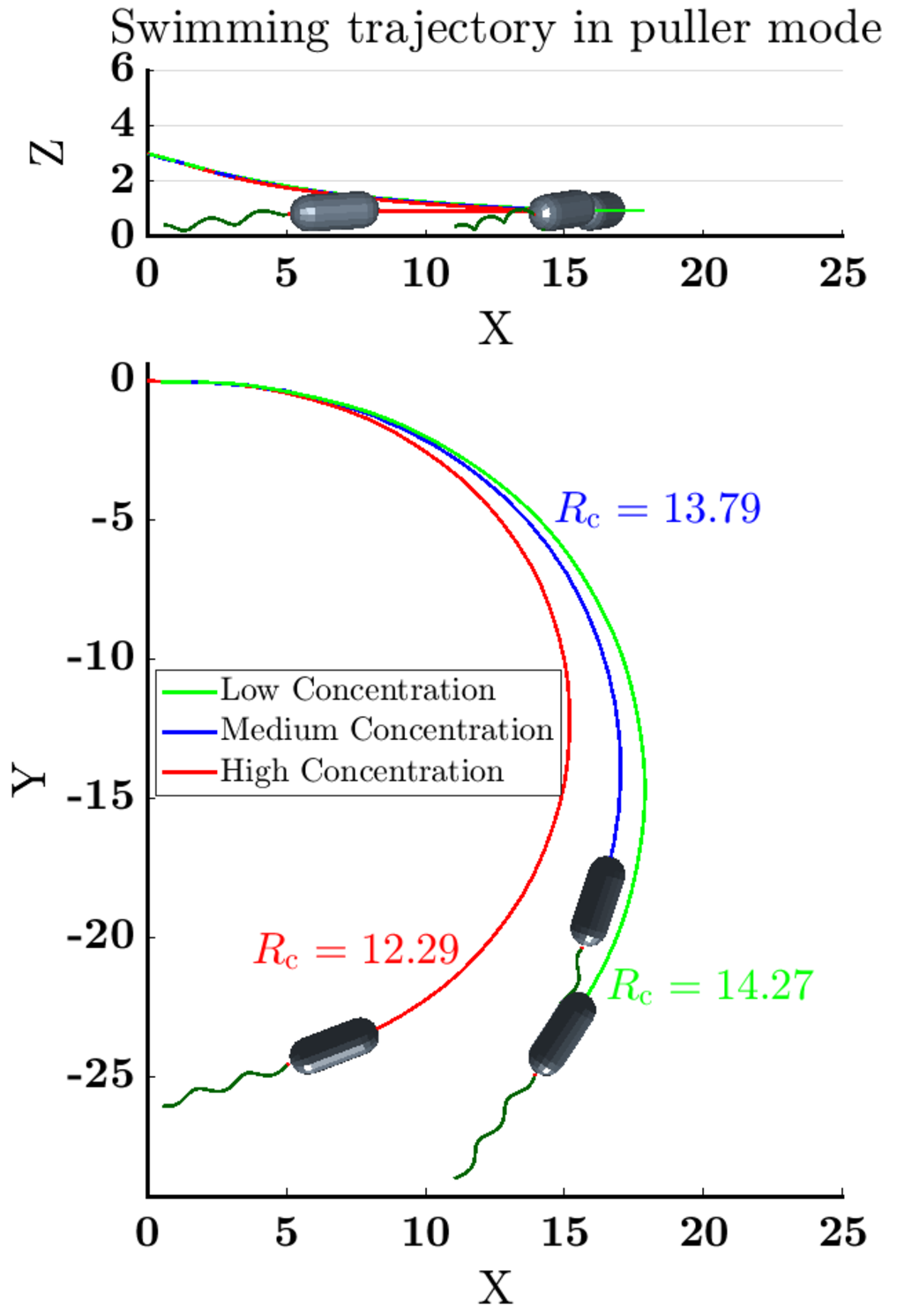}
\caption{\label{fig:PullerNearWall} Near surface swimming trajectory of the model bacterium in the puller mode. In different concentrations of NaCl, the model bacterium is entrapped by the surface. The NaCl concentration in the swimming medium changes the radius of the circular trajectories ($R_\mathrm{c}$). The initial conditions are the same as the pusher mode. Simulation times are $T_\mathrm{s}=17000$, $T_\mathrm{s}=15000$, and $T_\mathrm{s}=6800$ for the low, medium, and high concentrations, respectively.}
\end{figure}

As displayed in Fig.~\ref{fig:PullerNearWall}, the model bacterium in the puller mode is attracted to the surface, regardless of the Na$^+$ concentration. Our numerical results show that the model bacterium moves on orbits of higher curvature at higher concentrations of NaCl. The cell body is almost parallel to the surface in all cases but becomes more parallel at higher concentrations; the mean angle between the cell body's long axis and the surface at high, medium, and low concentrations of NaCl are $2.66^\circ$, $3.32^\circ$ and $4.11^\circ$, respectively. This trend with NaCl concentration is consistent with the hydrodynamic effects of the force dipole image system in the boundary. For pullers, the boundary-induced velocity pushes the hook away from the wall, reducing the inclination of the cell body away from the wall. 

\begin{figure}
\includegraphics[width=75mm]{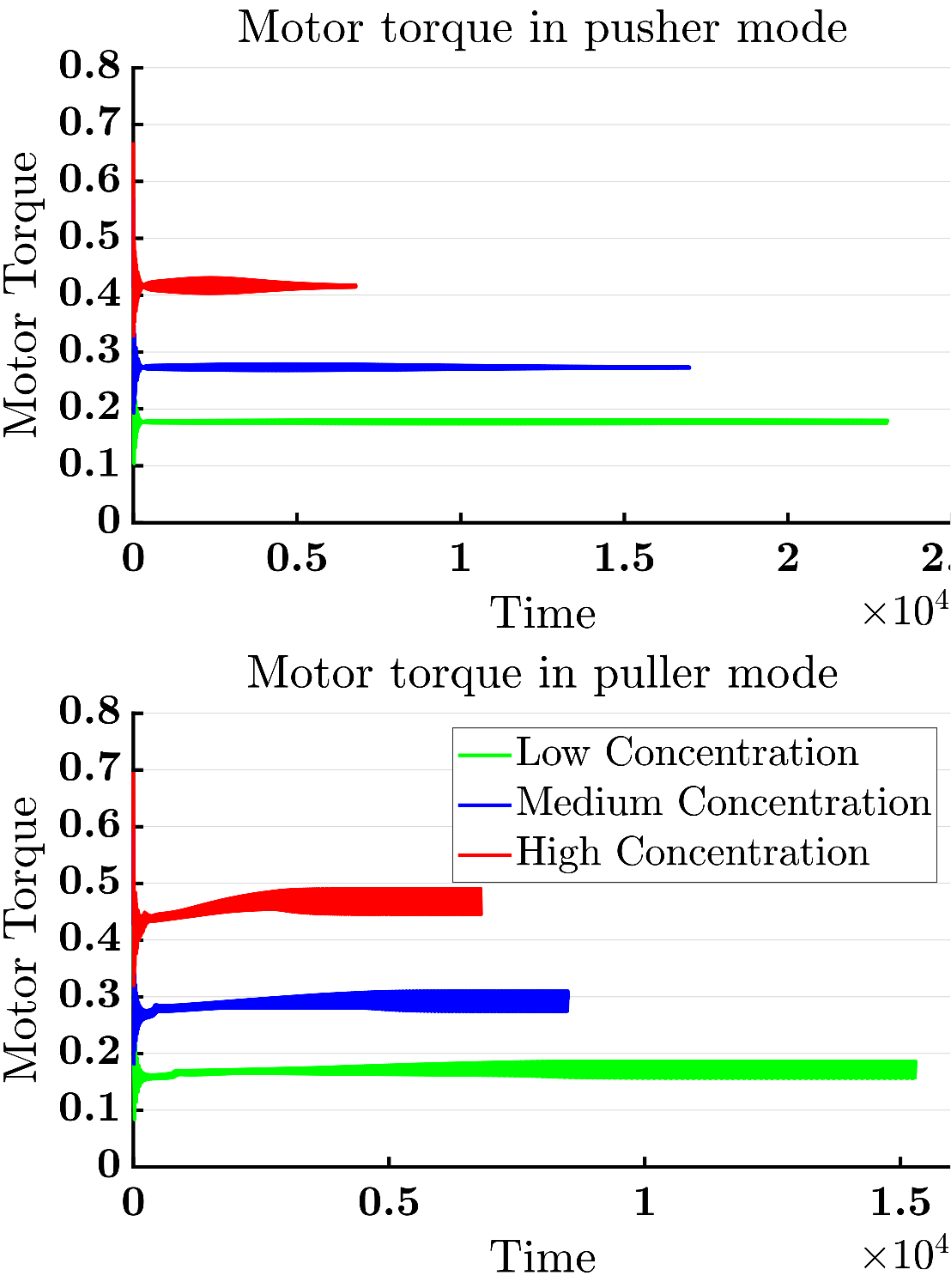}
\caption{\label{fig:MotorTorques} Variation of the motor load with time as the bacterium swims toward the surface and then is deflected away (pusher mode) or entrapped (puller mode)  by the surface.}
\end{figure}

Fig.~\ref{fig:MotorTorques} represents the variation of the motor torque as the model bacterium swims next to the surface. After a brief transition period from the initial condition to a quasi-steady swimming configuration, the mean value of the motor torque does not change significantly as the model bacterium swims toward or escapes from the surface. However, the motor torque oscillates with each revolution of the flagellum due to variations in the hydrodynamic and steric interactions with the wall. The amplitude of the motor torque oscillations is largest when the flagellum is close to the surface. In all cases the mean motor torque is higher in the puller mode, which indicates that the motor speed is lower in the puller mode. This is consistent with the results discussed for free space swimming in Section~\ref{subsec:swimmingspeed} and is enhanced by the fact that pullers tend to swim closer to the surface, where the increased drag reduces the motor speed.

\subsubsection{Sensitivity to initial conditions}

\begin{figure}
\includegraphics[width=88mm]{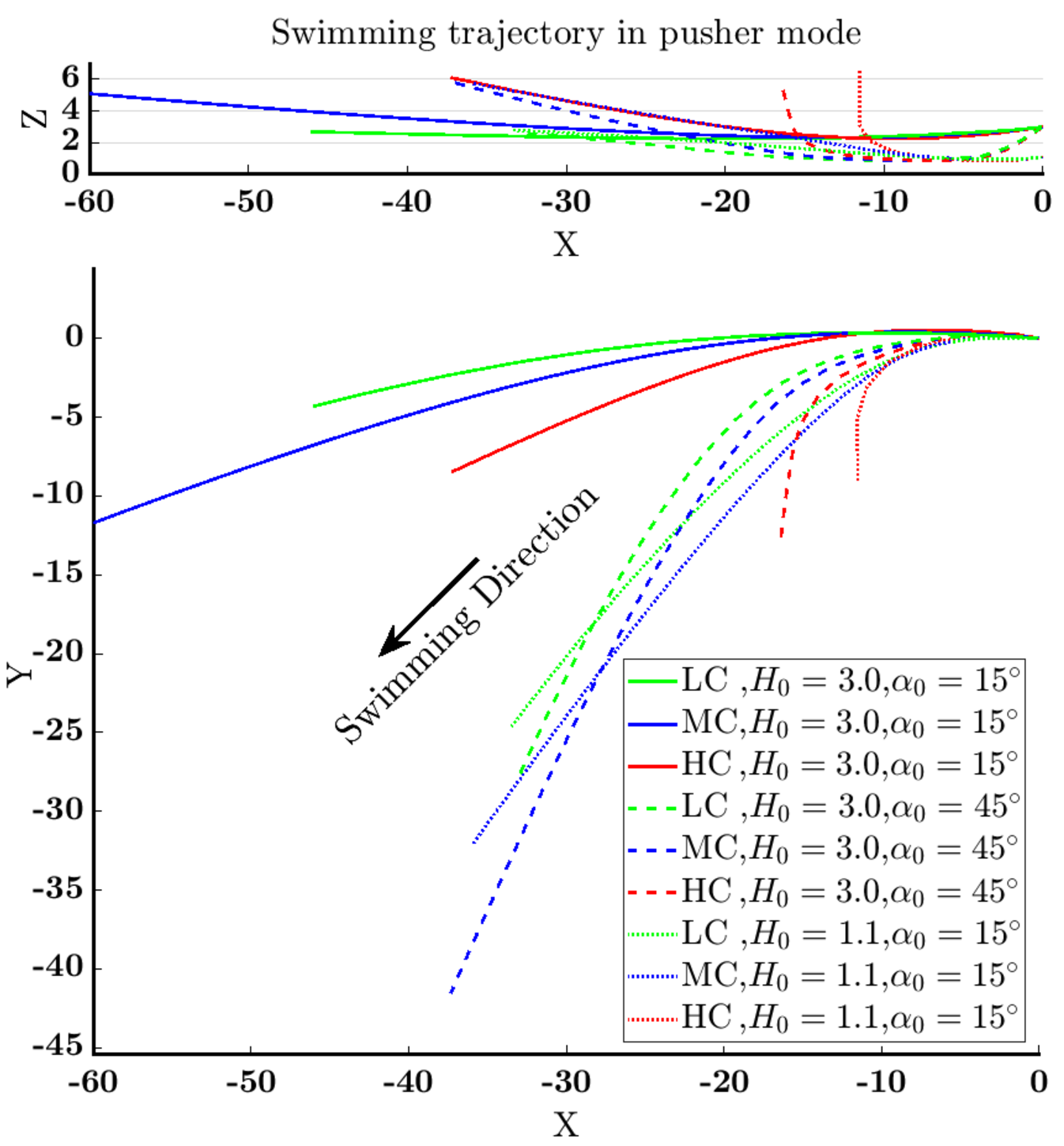}
\caption{\label{fig:Initialcondition} Swimming trajectories of the pusher-mode model bacterium in two different initial distances ($H_0=1.1,3$) from the surface and two attack angles ($\alpha_0=15^\circ$, $\alpha_0=45^\circ$). LC, MC, and HC are respectively abbreviations for low, medium, and high concentrations of NaCl.}
\end{figure}

We showed that the escaping angle of the pusher-mode model bacterium varies with the concentration of sodium chloride. To ensure that this near-surface behavior is independent of the initial distance and orientation, we compare the swimming trajectories of the model bacterium when it is initially placed in different distances ($H_0=1.1,3$) and angles ($\alpha_0=15^\circ,45^\circ$) from the surface. As shown in Fig.~\ref{fig:Initialcondition}, the model bacteria mainly remain near the surface in the lowest concentration of NaCl (green trajectories), and conversely, they exhibit a weak tendency to remain longer near the surface in the highest concentration of NaCl (see red trajectories).  The escaping angles are quantitatively compared in Tab.~\ref{tab:InitialCondition}. These results clearly illustrate that regardless of the initial condition, pusher-mode bacteria strongly tend to swim close to the surface in the lower concentrations of the ions. Such a correlation between the concentration and the tendency to mainly move next to the surface is consistent with the experimental observations of Wu et al.~\cite{wu2018}. The obtained results in Tab.~\ref{tab:InitialCondition} also demonstrate a meaningful correlation between the escaping angle and the initial distance and attack angle. In this respect, the bacterium escapes from the surface with a larger angle as it is initially placed closer to the surface and/or approaches the surface with a larger attack angle. Comparing the obtained angles indicate that the dependency of the escaping angle to the initial condition is notable in the high concentration of NaCl, and it is fairly negligible in the medium and low concentrations. 
\begin{table}[b]
\caption{\label{tab:InitialCondition}%
Escaping angles of the pusher-mode bacterium in different initial conditions and NaCl concentrations.}
\begin{ruledtabular}
\begin{tabular}{c|cccc}
\textbf{Con.} & $H_0=3$ & $H_0=3$ & $H_0=1.5$& $H_0=1.1$\\
              & $\alpha_{e}=45^\circ$ & $\alpha_{e}=15^\circ$ & $\alpha_{e}=15^\circ$ & $\alpha_{e}=15^\circ$\\
\hline
Low    & $1.65^\circ$  & $1.24^\circ$  & $1.55^\circ$ &  $1.74^\circ$\\
Medium & $6.19^\circ$  & $4.80^\circ$  & $5.76^\circ$ &  $6.17^\circ$\\
High   & $28.12^\circ$ & $10.98^\circ$ & $18.04^\circ$ &  $42.81^\circ$\\
\end{tabular}
\end{ruledtabular}
\end{table}%

\subsubsection{Effects of hook and filament flexibility}
\begin{figure}
\includegraphics[width=85mm]{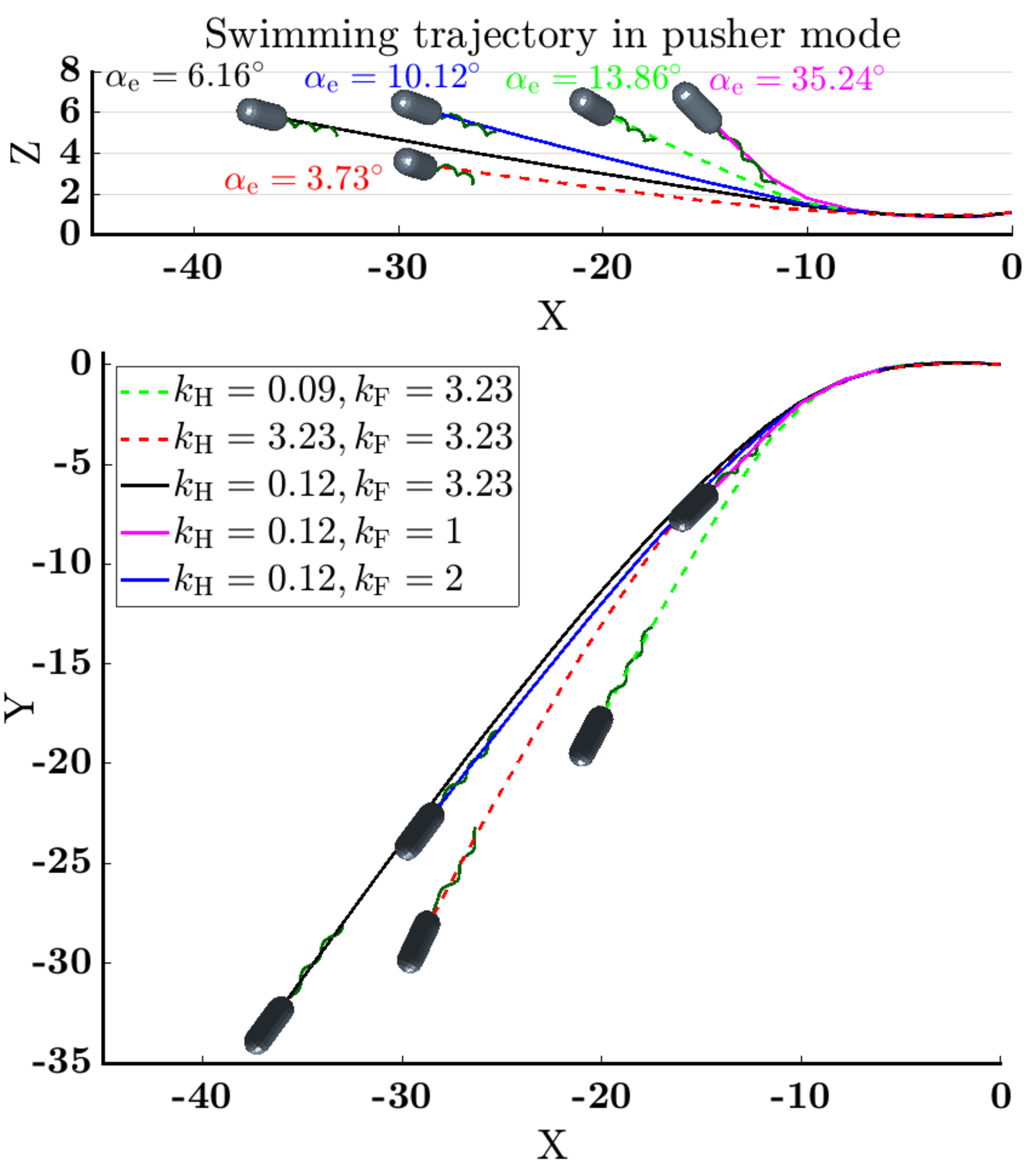}
\caption{\label{fig:PusherStiffness} Swimming trajectories of the pusher-mode model bacterium in the different flagellum and hook stiffnesses. The concentration of NaCl is medium in all the cases. A video corresponding to two of the trajectories is available in the Supplemental Material.}
\end{figure}

The hydrodynamic interactions between the uniflagellated bacteria and a planar surface have already been studied well when the flagellum is assumed to be a single rigid helix~\cite{ramia1993role,lauga2006swimming,shum2010modelling}. This simplification is adopted in many studies but the flexibility of the hook and the flagellum could impact attributes such as the mean swimming speed, the orientation of the cell body, and the flagellum with respect to the surface. Therefore, hook and flagellum flexibility could change the boundary accumulating behavior of the bacteria. To fill the research gap and better understand the behavior of different flagellated microorganisms near the surfaces, we study the locomotion of the model bacterium near the surface when different stiffnesses are assigned to the flagellum filament and the hook. As shown in Fig.~\ref{fig:PusherStiffness}, decreasing the rigidity of the filament and/or the hook helps the pusher-mode model bacterium to escape from the surface more easily. When the cell body is pushed near the surface, the viscous torque tends to tilt the cell body upwards. If the filament and hook are stiff, then the cell body cannot rotate because the flagellum is obstructed by the wall. When the filament or hook is more flexible, they bend easily and allow the cell body to rotate away from the wall and escape.

\begin{figure}
\centering
\includegraphics[width=85mm]{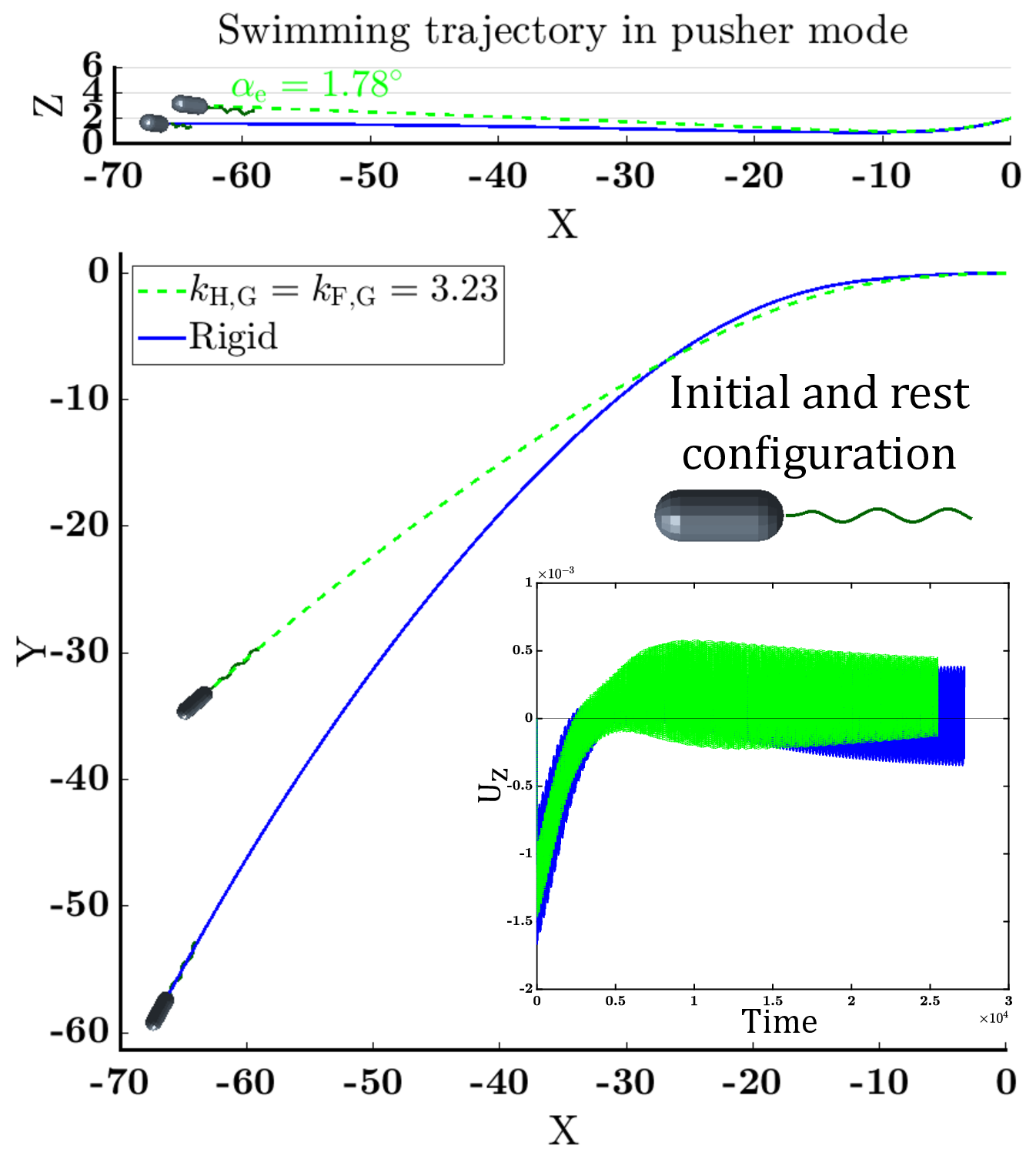}
\caption{\label{fig:RigidFlexible} The uniflagellated model bacterium with rigid flagellum is entrapped by the surface, whereas it is pushed back as the flagellum is flexible. The amplitude growing factor $k_\mathrm{E}=1.83$ is used to describe the flagellum shape and align the flagellum and cell body axes. There is no hook and the entire filament's relative stiffness is $k_\mathrm{F}=3.23$. The concentration of NaCl is medium. Note that the subscript G in the legend indicates that the flagellum shape with a growing factor was used.}
\end{figure}

\begin{figure}
\includegraphics[width=75mm]{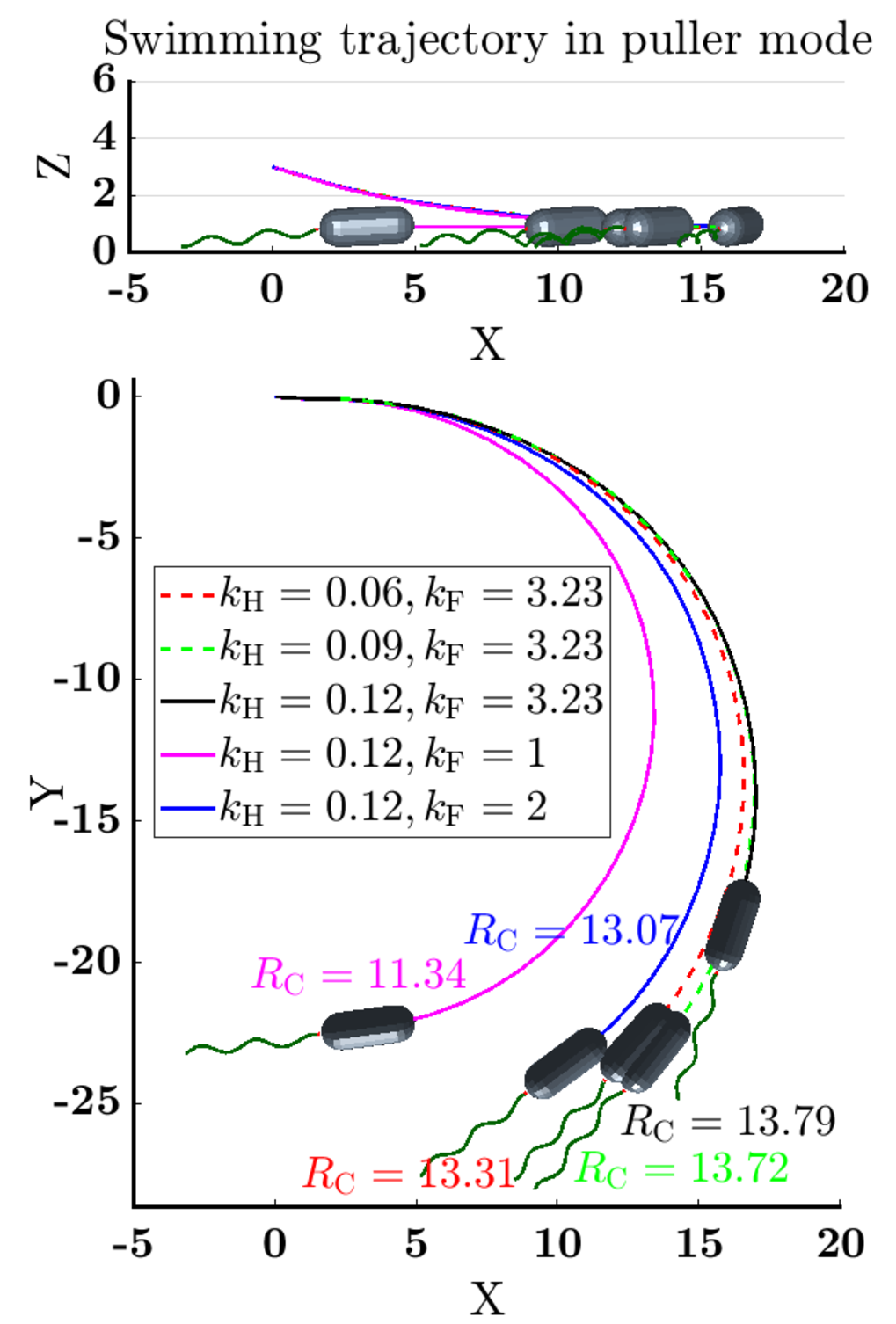}
\caption{\label{fig:PullerStiffness} Swimming trajectories of the puller-mode model bacterium in different flagellum and hook stiffnesses. The concentration of NaCl is medium in all cases. A video corresponding to one of the trajectories is available in the Supplemental Material.}
\end{figure}

Motivated by the obtained results which indicate that the escaping angle decreases by increasing the flagellum or hook rigidity, we compare the behavior of two model bacteria with a flexible and rigid flagellum, respectively. For both models, we define the flagellum helical shape with an amplitude envelope $k_\mathrm{E}$ to align the flagellum with the cell body axes. This shape is used instead of the purely helical filament so that the axis of the rigid flagellum is aligned with the axis of the cell body, as required for effective propulsion. For the flexible model, we use the stiffness $k_\mathrm{F} = 3.23$ and do not consider a separate hook structure. Interestingly, our simulations (Fig.~\ref{fig:RigidFlexible}) show that the model bacterium with rigid flagellum is entrapped by the surface whereas the bacterium with a flexible flagellum escapes from the surface with a small escaping angle. The simulations are continued beyond the trajectories shown to ensure that the bacteria escape or remain at the surface as described. These simulations demonstrate that the flexibility of the flagellum in pusher-mode bacteria likely facilitates the escape from the surfaces. 
Comparing the trajectories in the puller mode (Fig.~\ref{fig:PullerStiffness}) shows that the model bacterium moves on smaller circular orbits when it has a more flexible hook or flagellum. Furthermore, the radius of the orbits mainly changes with the filament stiffness than the hook stiffness. Calculating the stable orientation of the cell body demonstrates that the long axis of the cell body is more parallel to the surface as the hook or the filament is stiffer. In our simulations, this angle varies from $6.4^\circ$ to $3.3^\circ$, depending on the stiffnesses. Like the pusher mode, the cell body more freely tilts upwards when the filament and hook are more flexible, hence the cell body makes a larger angle with respect to the surface. We note that the sensitivity of the path curvature on flagellar stiffness contrasts with results by Park et al.~\cite{park2019flagellated}, who found that the circular orbits of pusher bacteria were unaffected by the prescribed motor frequency (which should have the same effect as varying the stiffness of the flagellum).  

\subsubsection{Hook instability}

The hook in uniflagellated bacteria is very flexible and easily becomes unstable (buckled) if it is subjected to a load which is more than a critical value. When bacteria swim toward a boundary, the load on the hook deviates from the steady state load in unbounded fluid due to hydrodynamic interactions with the boundary. These fluctuations in viscous forces could make the hook much more susceptible to becoming unstable.

In this section, we prescribe a constant motor torque $T=1$ and choose the hook's stiffness so that the hook is stable but close to the free-space critical value for pushers, i.e., the threshold rigidity below which the hook becomes unstable in an unbounded fluid. By performing simulations with different hook stiffnesses, the critical rigidity of the hook is determined to be $k_\mathrm{H}\approx0.105$. We vary the hook rigidity starting at a minimum value $k_\mathrm{H}=0.106$ and study the locomotion of the pusher mode model bacterium near a surface. The other model parameters are as listed in Tab.~\ref{tab:physicalparameters}.

\begin{figure}
\centering
\includegraphics[width=80mm]{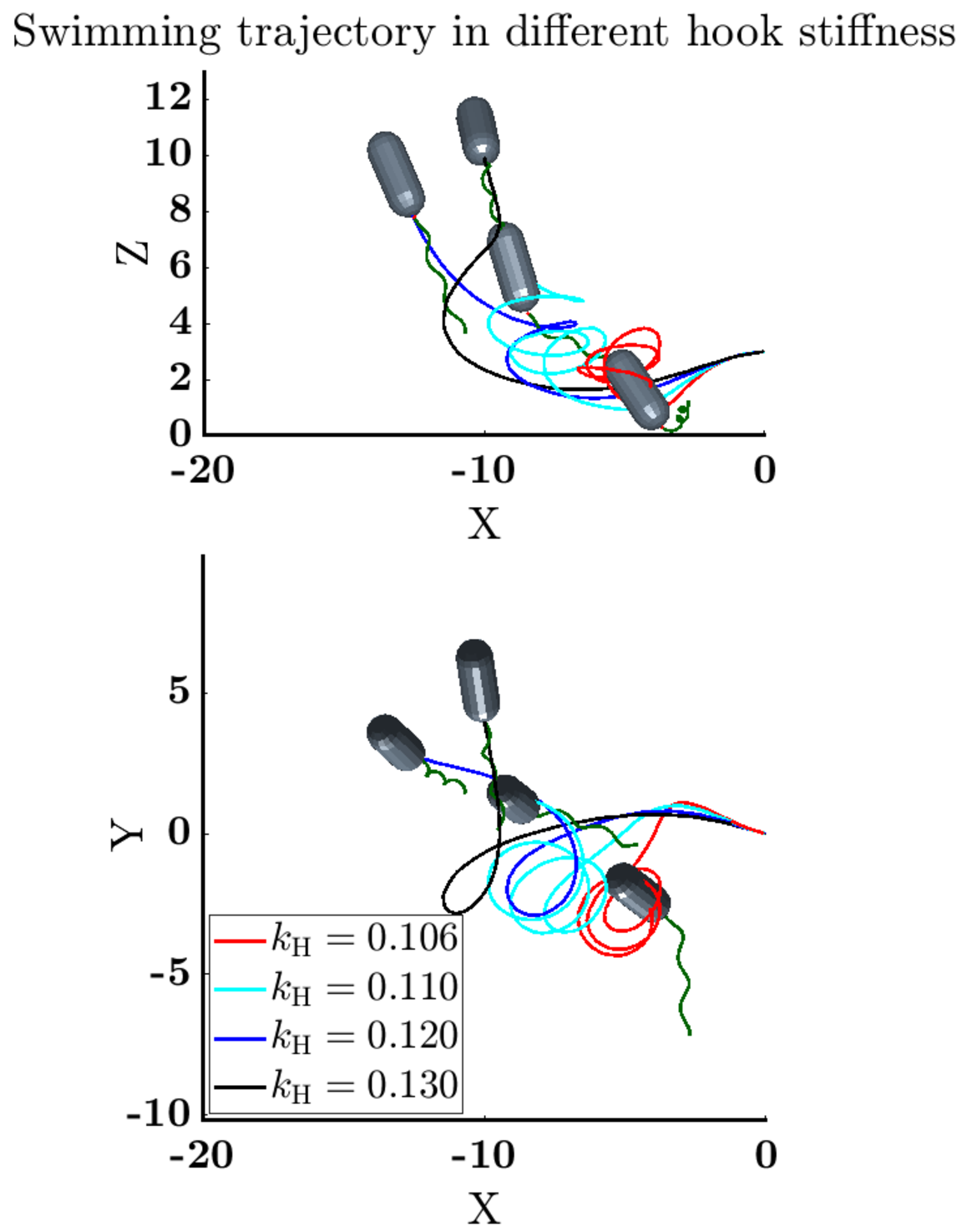}
\caption{\label{fig:HookInstability} Near-surface trajectories of bacteria with relative hook stiffnesses close to the free-space threshold for instabilities. High viscous forces near a planar surface increase the load applied to the hook and make it unstable. As the bacterium moves away from the surface, the hook becomes stable and the bacterium travels on a straight trajectory. When the hook's relative stiffness is slightly above its critical value ($k_\mathrm{H}=0.106$), the bacterium is unable to escape from the surface. A video corresponding to two of the trajectories is available in the Supplemental Material.}
\end{figure}

As shown in Fig.~\ref{fig:HookInstability}, the trajectories are distinct from the gradually escaping paths and entrapped circular orbits typically observed. Instead, the cell body undergoes rapid reorientation and swims away from the wall with a large escape angle (for $k_\mathrm{H} \geq 0.110$), or even becomes trapped with the bent hook against the wall and both the cell body and the flagellum pointing away from the wall (for $k_\mathrm{H} = 0.106$). In the latter case, the cell body spins and the flagellum precesses around the cell body but there is no appreciable net swimming motion either away from or parallel to the wall. The large deformations of the hook indicate that the load on the hook exceeds the critical value for buckling when the swimmer is near a wall. If the relative hook stiffness is high enough, the hook only buckles transiently before returning to a stable swimming shape but if the relative hook stiffness is low, then the hook does not recover from the buckled state. Interestingly, the spinning motion at $k_\mathrm{H} = 0.106$ is reminiscent of experimental observations of \textit{V.~fischeri} intermittently pausing, wiggling, and changing directions when confined between parallel plates~\cite{tokarova2021patterns}. Also, the abrupt changes in direction in the cases with higher relative hook stiffness are similar to the flick behavior of \textit{V.~alginolyticus}, which is due to a dynamic instability that occurs when the motor switches from reverse to forward swimming~\cite{son2013,jabbarzadeh_dynamic_2018}. The hook instability near walls that we see in our simulations could provide a mechanism for tumbling without any change in motor torque or direction.

\subsubsection{Effects of cell body aspect ratio}
\begin{figure}
\includegraphics[width=88mm]{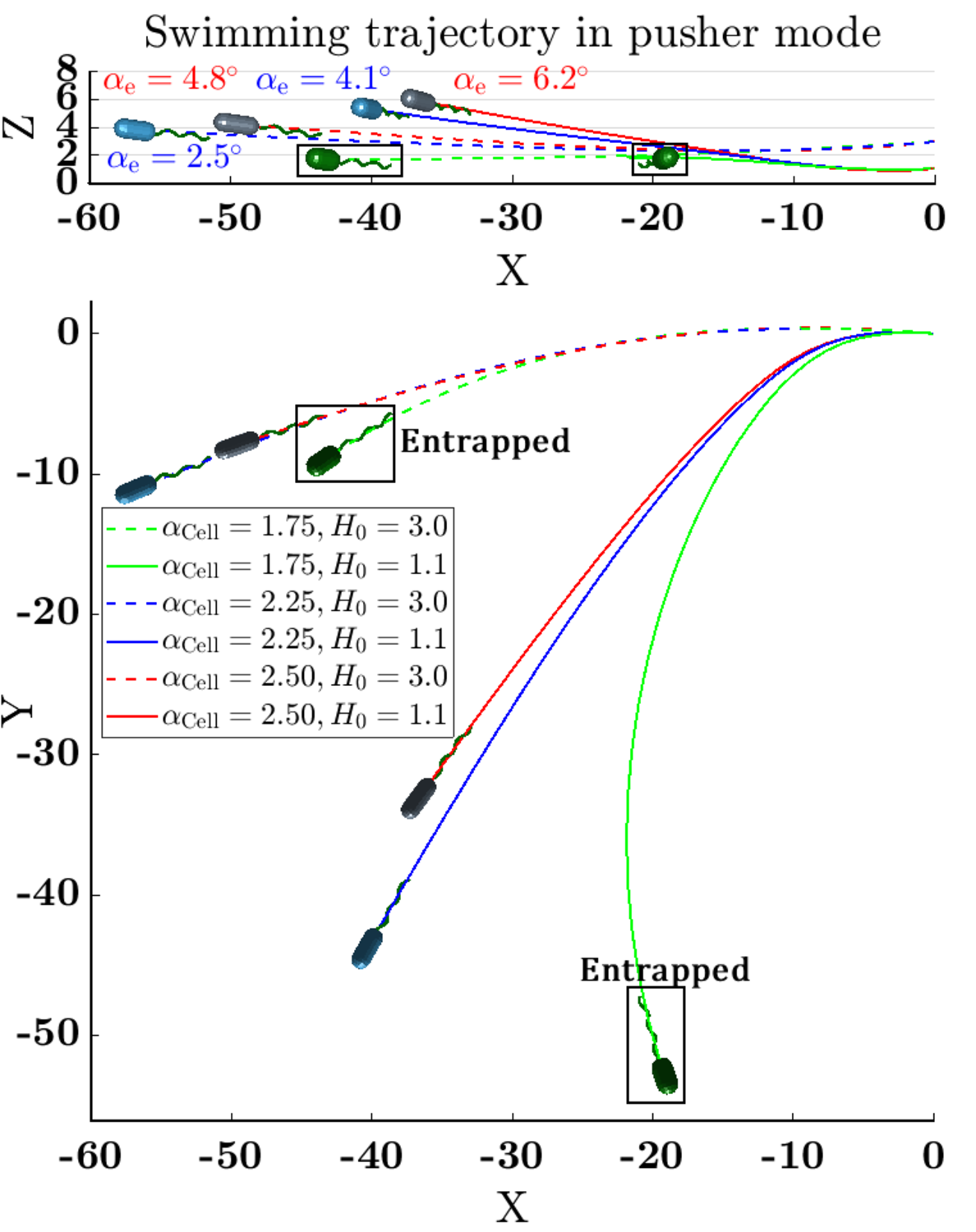}
\caption{\label{fig:AspectRatio}Swimming trajectories of the pusher-mode model bacterium with different cell body aspect ratios. The bacteria with the largest cell body's aspect ratio, $\alpha_\mathrm{Cell}=2.5$, escape from the surface with relatively large angles. In the smallest aspect ratio, $\alpha_\mathrm{Cell}=1.75$, the bacteria are entrapped near the surface. The concentration of NaCl is medium in all cases.}
\end{figure}

Our results thus far have shown that the model pusher bacterium (with aspect ratio $\alpha_\mathrm{cell}=2.5$) escapes from the surface regardless of the different flagellum stiffness, NaCl concentrations, and the initial conditions chosen in this study. Previous numerical investigations have shown that decreasing the aspect ratio of the cell body increases the tendency of pusher-mode bacteria to be hydrodynamically trapped near the surfaces~\cite{shum2010modelling, park2019flagellated}. To illustrate the importance of the cell body aspect ratio in the entrapment of pusher-mode bacteria, we reduce the cell body's aspect ratio from 2.5 to 2.25 and 1.75. As expected and shown in Fig.~\ref{fig:AspectRatio}, the escaping angle of the model bacterium decreases when the aspect ratio becomes 2.25. Further reduction of the aspect ratio (to $\alpha_\mathrm{cell}=1.75$), causes the bacterium to be entrapped at the surface. Our simulations demonstrate that independent of the initial distance from the surface, the bacterium reaches a unique stable distance $H_\mathrm{c}=1.71$ from the surface when the aspect ratio is  $\alpha_\mathrm{cell}=1.75$.

\begin{figure}
\centering
\includegraphics[width=90mm]{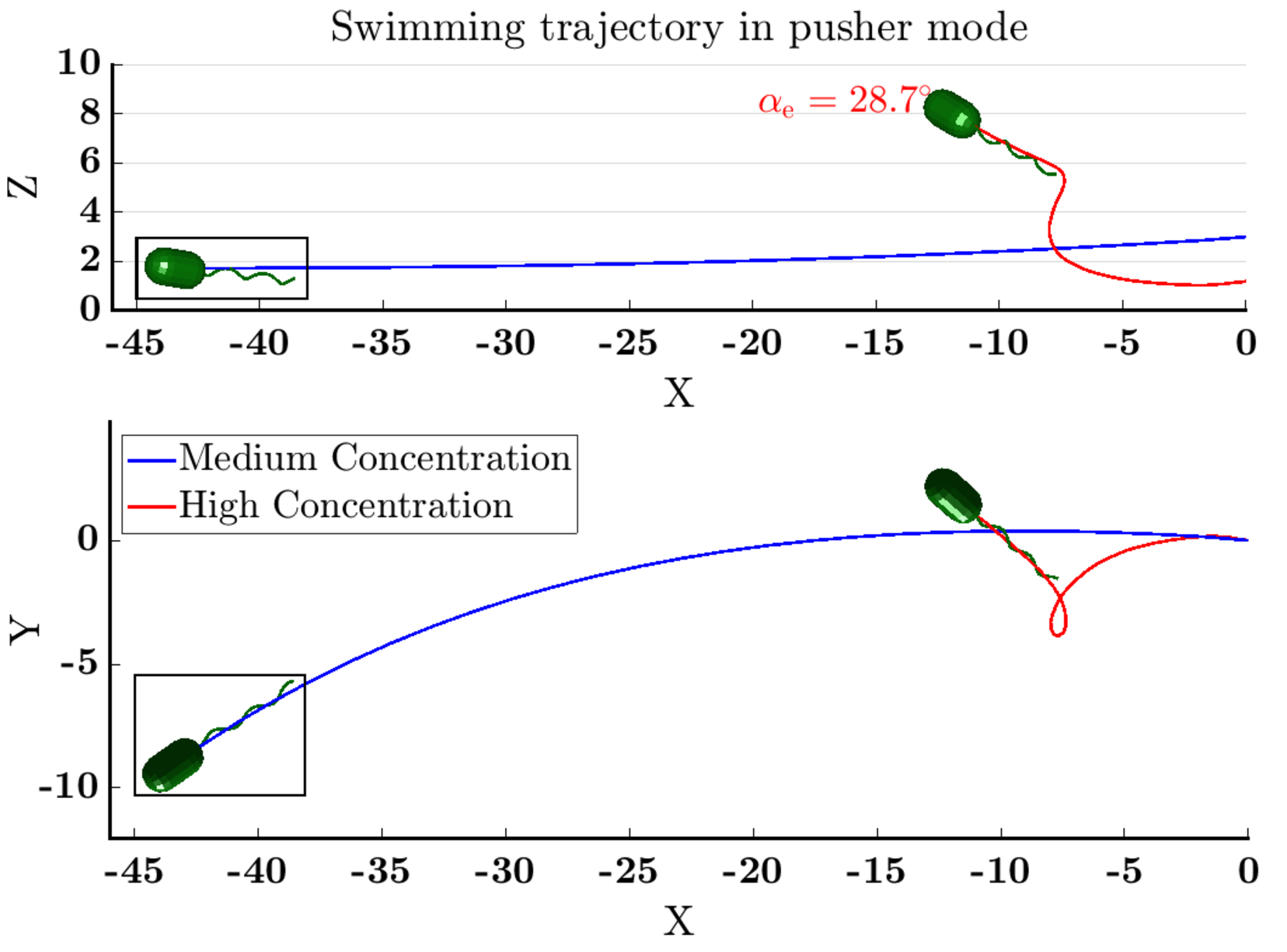}
\caption{\label{fig:OneHalfSpeeds} Changing the ions concentration from medium to high causes the boundary accumulating pusher-mode bacteria to escape from the boundary. In these simulations, $\alpha_{cell}=1.75$, and the other parameters are stated in Tab.~\ref{tab:physicalparameters}. A video corresponding to these trajectories is available in the Supplemental Material.}
\end{figure}

Having found that for $\alpha_\mathrm{cell}=1.75$, the model bacterium with flexible flagellum and hook is attracted to the surface, we next consider the motion under higher concentrations of NaCl to see whether this qualitatively affects the behavior near boundaries, Interestingly, the model bacterium escapes from the surface when the concentration increases from medium to high, as shown in Fig.~\ref{fig:OneHalfSpeeds}. This result is consistent with experimental evidence that the concentration of ions changes the pusher-mode bacteria's behavior in boundary accumulating; specifically, they tend to escape from the surfaces at higher concentrations of NaCl.

\section{Discussion and conclusion}

The main aim of this study is to model and analyze the near-surface motion of uni-flagellated bacteria with a flexible hook and filament using parameters appropriate for \emph{V.~alginolyticus}. Unlike other modelling studies~\cite{park2019flagellated,shum2012effects,park2019locomotion}, we employ empirical relationships between the flagellar motor torque and its frequency to account for the load-dependent motor activity. Using a straight hook and purely helical filament shape, we show that the swimming speed for a fixed motor torque does not change significantly with the flagellum stiffness (over the tested range). Applying the characteristic motor torque--speed relationship, however, the motor torque increases with flagellar stiffness for pushers and has the opposite trend for pullers; these effects are most pronounced at high NaCl concentrations. This suggests that changes in the propulsive efficiency of the flagellum due to deformations are compensated by changes in the motor torque and speed when the flagellum is deformed. The implication is that a model that assumes constant torque would predict no change in swimming speed with stiffness whereas a model that assumes constant motor speed would predict higher swimming speeds for pushers and lower swimming speeds for pullers at higher flagellar stiffnesses. Accounting for the motor torque--speed relationship, the trend with flagellar stiffness is as in the constant motor speed model but to a lesser degree.

At medium NaCl concentrations and high flagellar stiffnesses, we found that the motor torque and frequency are approximately the same for pushers and pullers but pushers swim around 8\% faster than pullers. This trend is sensitive to the model used for the hook and filament shape, however. If the flagellum shape is instead defined with a growing amplitude, then pullers are around 3\% faster than pushers, which is similar to the findings of Park et al.~\cite{park2019locomotion} using constant motor speed and the same growing amplitude model for the flagellum. 


Experimental observations have shown that the accumulation of \emph{V.~alginolyticus} near surfaces changes with the concentration of sodium chloride in the swimming medium. Depending on the swimming mode (pusher or puller), the relationship between the ion concentration and the tendency to swim near the surfaces could be direct or inverse~\cite{wu2018}. We confirm that changing the ion concentration, assuming this only shifts the motor torque--speed curve, affects the near-surface behavior of bacteria.  In particular, for certain geometries and mechanical properties of the model bacterium, the pusher mode swimmer is attracted to surfaces at low ion concentrations and escapes at high concentrations. In this regard, comparing the escaping angles of the pusher-mode model bacterium in different concentrations of NaCl shows an inverse relationship between these parameters. Further investigation indicates that this conclusion is independent of the initial conditions of the bacteria.

Our results in the puller mode show that for a model bacterium that is attracted to surfaces at all concentrations of NaCl, variations in the ion concentration impacts the radius of the circular orbits and the stable orientation of the cell body with respect to the surface. In particular, the model bacterium tends to move on smaller circular paths in higher concentrations of NaCl. Our simulations of \emph{V.~alginolyticus} swimming in puller mode exhibit circular orbits of radius $R_\mathrm{c}\approx$17.5\,\textmu{}m when converted to dimensional units; this is comparable to the experimental measurements ($R_\mathrm{c}\approx$10-15.5\,\textmu{}m) of Wu et~al.~\cite{wu2018}. 

In addition to \emph{V.~alginolyticus}, the uniflagellated bacterium \textit{Caulobacter crescentus} has been experimentally observed swimming forwards and backwards near surfaces~\cite{li2011accumulation}. It was reported that this bacterium spends much less time close to the surface when swimming in pusher mode compared with swimming in puller mode. Circular orbits were only observed in backward swimming cells. These observations are consistent with our numerical results using parameters for \textit{V.~alginolyticus}.

We note that the flexibility of the hook and filament (as long as they are in a stable state) facilitates the escaping from the surface by allowing the cell body to tilt upward more freely. Our simulations illustrate that the flexibility of the flagellum may change a pusher-mode model bacterium state from boundary accumulating to boundary escaping, for example. In general, it seems that there is an inverse relationship between the cell body's long axis angle with the surface and the hook or filament's relative stiffness in either puller or pusher modes.

Higher viscous forces applied to the flagellum and the cell body as the bacterium swims near a surface may cause the hook to become unstable. This kind of instability may lead to a different form of entrapment near surfaces in which the cell body spins on the spot with the hook bent and closest to the surface. Transient instability of the hook due to proximity to a surface results in an abrupt and large change in orientation of the cell, similar to a flick. These two passive behaviors, separated by a small change in relative hook stiffness, have opposite consequences; the former case keeps the cell pressed against the surface whereas the latter quickly scatters the cell away from the surface. Interestingly, it has been reported that tumbling is suppressed near solid surfaces for the peritrichous bacterium \textit{Escherichia coli}~\cite{molaei_failed_2014}. The effects of surfaces on bacteria are, evidently, highly dependent on the mode of motility.  

The simulations show that the pusher-mode bacterium with a flexible (but far from the threshold for instabilities) hook and filament is entrapped by the flat surface when the cell body has a small aspect ratio. This conclusion is consistent with the results of Park et al.~\cite{park2019flagellated}.  The transition from escaping state to entrapment state in a specific aspect ratio of the cell body is already well studied for the bacteria with rigid flagellum~\cite{shum2010modelling}. Here, comparing the near-surface behavior of a rigid flagellum model with that of a flexible flagellum model demonstrates that the flexibility of the flagellum can affect the threshold of the cell body aspect ratio for surface entrapment. 

To sum up, whereas many investigations of uniflagellated bacterial locomotion are carried assume that the bacterial flagellum is rigid, our results clearly demonstrate that the hook and flagellum flexibility may change the behavior of the bacterium, especially near a planar surface.  For example, flexibility may change the bacteria's behavior from boundary accumulating to boundary escaping or cause them to be locally entrapped near the surfaces. We expect that accurately accounting for hook and filament flexibility is also necessary for modelling bacteria interacting with each other and in other confined geometries.

\begin{acknowledgments}
We acknowledge the support of the Natural Sciences and Engineering Research Council of Canada (NSERC), [funding reference number RGPIN-2018-04418].

Cette recherche a \'{e}t\'{e} financ\'{e}e par le Conseil de recherches en sciences naturelles et en g\'{e}nie du Canada (CRSNG), [num\'{e}ro de r\'{e}f\'{e}rence RGPIN-2018-04418].
\end{acknowledgments}

\appendix

\section{Regularized stokeslet and rotlet}\label{Regularized stokeslet}
Consider a regularized point force $\vec{f}$ and torque $\vec{n}$ applied at $\vec{X} = (X,Y,h)$ and the associated image point $\vec{\hat{X}} = (X,Y,-h)$ due to a no-slip wall at $Z=0$. Defining the displacement vectors $\vec{r}=\vec{x}-\vec{X}$ and $\vec{\hat{r}}=\vec{x}-\vec{\hat{X}}$ to the evaluation point $\vec{x}$, the translational and angular velocities of the
regularized stokeslet and rotlet used in Eqs. (\ref{equ:BIET}) and (\ref{equ:BIEA}) are given by~\cite{park2019flagellated}:
{\small\begin{align}
\vec{U}_s&(\vec{f},\vec{r},\vec{\hat{r}},\epsilon)=\frac{1}{8\pi\mu}\bigg\{\Big[\vec{f} J_1(r,\epsilon)+\big(\vec{f}\cdot \vec{r}\big)\vec{r} J_2(r,\epsilon)\Big]\nonumber\\
&-\Big[\vec{f} J_1(\hat{r},\epsilon)+\big(\vec{f}\cdot \vec{\hat{r}}\big)\vec{\hat{r}} J_2(\hat{r},\epsilon)\Big]
 -h^2\Big[\big(\vec{b} \cdot \vec{\hat{r}}\big)\vec{\hat{r}} K_2(\hat{r},\epsilon)\nonumber\\
 &+\vec{b}\,K_1(\hat{r},\epsilon)\Big]+2h\bigg[\big(\vec{b}\cdot \vec{e}_3\big)\vec{\hat{r}} J_2(\hat{r},\epsilon)
+\big(\vec{\hat{r}}\cdot\vec{b}\big)\vec{e}_3\big(J_3(\hat{r},\epsilon)\nonumber\\
&-J_2(\hat{r},\epsilon)\big)+\big(\vec{\hat{r}} \cdot \vec{e}_3\big)\vec{b} J_2(\hat{r},\epsilon)+\frac{1}{2}\big(\vec{\hat{r}} \cdot \vec{e}_3\big)\big(\vec{\hat{r}} \cdot \vec{b}\big)\vec{\hat{r}} K_2(\hat{r},\epsilon)\bigg]\nonumber\\
&+2h J_3(\hat{r},\epsilon)\big(\vec{m} \times \vec{\hat{r}}\big)\bigg\},
 \end{align}}%
 {\small\begin{align}
 \vec{U}_r&(\vec{n},\vec{r},\vec{\hat{r}},\epsilon)=\frac{1}{8\pi\mu}\bigg\{\frac{1}{2}\Big[P(r,\epsilon)\big(\vec{n} \times \vec{r}\big)- P(\hat{r},\epsilon)\big(\vec{n} \times \vec{\hat{r}}\big)\Big]\nonumber\\
 &+h\Big[\vec{p} K_1(\hat{r},\epsilon)+\big(\vec{p} \cdot \vec{\hat{r}}\big)\vec{\hat{r}}  K_2(\hat{r},\epsilon) \Big]-\bigg[\Big[\big(\vec{p} \cdot \vec{\hat{r}}\big)\vec{e}_3\nonumber\\
 &+\big(\vec{e}_3 \cdot \vec{\hat{r}}\big)\vec{p}\Big]J_3(\hat{r},\epsilon)+\big(\vec{e}_3 \cdot \vec{\hat{r}}\big)\big(\vec{p} \cdot \vec{\hat{r}}\big)\vec{\hat{r}}K_2(\hat{r},\epsilon)\bigg]\nonumber\\
&-J_3(\hat{r},\epsilon)\big(\vec{q} \times \vec{\hat{r}}\big)+h^2\,J_4(\hat{r},\epsilon)\big(\vec{n} \times \vec{\hat{r}}\big)
 -h\Big[\vec{p} J_3(\hat{r},\epsilon)\nonumber\\
 &+\big(\vec{e}_3 \cdot \vec{\hat{r}}\big)\big(\vec{n} \times \vec{\hat{r}}\big) J_4(\hat{r},\epsilon)\Big]\bigg\},
 \end{align}}%
 {\small\begin{align}
 \vec{W}_s&(\vec{f},\vec{r},\vec{\hat{r}},\epsilon)=\frac{1}{8\pi\mu}\bigg\{1/2\Big[P(r,\epsilon)\big(\vec{f} \times \vec{r}\big)- P(\hat{r},\epsilon)\big(\vec{f} \times \vec{\hat{r}}\big)\Big]\nonumber\\
 &+h^2 J_4(\hat{r},\epsilon)\big(\vec{b} \times \vec{\hat{r}}\big)+h\Big[K_2(\hat{r},\epsilon)-J_4(\hat{r},\epsilon)\Big]\big(\vec{b} \cdot \vec{\hat{r}}\big)\big(\vec{e}_3 \times \vec{\hat{r}}\big)\nonumber\\
 &+h\bigg[\vec{m}\Big[\hat{r}^2 J_4(\hat{r},\epsilon)+2J_3(\hat{r},\epsilon)\Big]-J_4(\hat{r},\epsilon)\big(\vec{m} \cdot \vec{\hat{r}}\big)\vec{\hat{r}}\bigg]\nonumber\\
 &+h P(\hat{r},\epsilon)\vec{m}\bigg\},
 \end{align}}%
{\small\begin{align}
\vec{W}_r&(\vec{n},\vec{r},\vec{\hat{r}},\epsilon)=\frac{1}{8\pi\mu}\bigg\{\frac{-1}{4}\Big[K_3(r,\epsilon)\vec{n}+K_4(r,\epsilon)\big(\vec{n}\cdot\vec{r}\big)\vec{r}\nonumber\\
&-K_3(\hat{r},\epsilon)\vec{n}-K_4(\hat{r},\epsilon)\big(\vec{n}\cdot\vec{\hat{r}}\big)\vec{\hat{r}}\Big]+\big(\vec{p} \cdot \vec{\hat{r}}\big)\big(\vec{e}_3 \times \vec{\hat{r}}\big)+\frac{1}{2}\Big[J_4(\hat{r},\epsilon)\nonumber\\
&-K_2(\hat{r},\epsilon)\Big]\big(\vec{e}_3 \cdot \vec{\hat{r}}\big)\big(\vec{p} \times \vec{\hat{r}}\big)-h J_4(\hat{r},\epsilon)\big(\vec{p} \times \vec{\hat{r}}\big)-\frac{1}{2}\Big[\hat{r}^2 J_4(\hat{r},\epsilon)\nonumber\\
 &+2J_3(\hat{r},\epsilon)\Big]\vec{q}+\frac{1}{2}J_4(\hat{r},\epsilon)\big(\vec{q} \cdot \vec{\hat{r}}\big)\vec{\hat{r}}-\frac{h}{2}\bigg[\big(\vec{e}_3 \cdot \vec{\hat{r}}\big)\vec{n}\Big[\hat{r}^2 J_5(\hat{r},\epsilon)\nonumber\\
 &+3J_4(\hat{r},\epsilon)\Big]-J_4(\hat{r},\epsilon)\Big[\big(\vec{n} \cdot \vec{e}_3\big)\vec{\hat{r}}+\big(\vec{p} \times \vec{\hat{r}}\big)\Big]\nonumber\\
 &-J_5(\hat{r},\epsilon)\big(\vec{e}_3 \cdot \vec{\hat{r}}\big)\big(\vec{n}\cdot \vec{\hat{r}}\big)\vec{\hat{r}}\bigg]+\frac{h^2}{2}\bigg[\vec{n}\Big[2J_4(\hat{r},\epsilon)+\hat{r}^2 J_5(\hat{r},\epsilon)\Big]\nonumber\\
 &-J_5(\hat{r})\big(\vec{n}\cdot \vec{\hat{r}}\big)\vec{\hat{r}}\bigg] \bigg\},
\end{align}}%
where
{\small\begin{align}
\vec{b}&=2(\vec{f} \cdot \vec{e}_3)\vec{e}_3 - \vec{f},\\
\vec{m}&=\vec{f} \times \vec{e}_3,\\
\vec{p}&=\vec{n} \times \vec{e}_3,\\
\vec{q}&=\vec{n} - (\vec{n} \cdot \vec{e}_3)\vec{e}_3,
\end{align}}%
{\small\begin{align}
J_1(r,\epsilon)&=\frac{2\epsilon^2+r^2}{ (r^2+\epsilon^2)^{3/2}},\label{equ:J1equation}\\
J_2(r,\epsilon)&=\frac{1}{(r^2+\epsilon^2)^{3/2}},\\
J_3(r,\epsilon)&=\frac{-3\epsilon^2}{(r^2+\epsilon^2)^{5/2}},\\
J_4(r,\epsilon)&=\frac{15\epsilon^2}{(r^2+\epsilon^2)^{7/2}},\\
J_5(r,\epsilon)&=\frac{-105\epsilon^2}{(r^2+\epsilon^2)^{9/2}},\\
P(r,\epsilon)&=\frac{5\epsilon^2+2r^2}{ (r^2+\epsilon^2)^{5/2}},\\
K_1(r,\epsilon)&=\frac{-10\epsilon^4+7\epsilon^2r^2+2r^4}{ (r^2+\epsilon^2)^{7/2}},\\
K_2(r,\epsilon)&=\frac{-21\epsilon^2-6r^2}{ (r^2+\epsilon^2)^{7/2}},\\
K_3(r,\epsilon)&=\frac{-4\epsilon^2+2r^2}{ (r^2+\epsilon^2)^{5/2}},\\
K_4(r,\epsilon)&=\frac{-6}{ (r^2+\epsilon^2)^{5/2}}.
\end{align}}

\nocite{*}
\providecommand{\noopsort}[1]{}\providecommand{\singleletter}[1]{#1}%

\end{document}